\documentclass[a4paper,11pt]{article}
\usepackage{jinstpub} 
\usepackage{lineno}

\usepackage{enumerate}
\usepackage{xcolor}
\usepackage{xspace}
\usepackage[acronym,toc]{glossaries}

\usepackage{cleveref}
\Crefname{figure}{Figure}{Figures} 
\crefname{figure}{figure}{figures} 
\crefname{equation}{eq.}{eqs.}

\usepackage{amsmath}
\usepackage{amsfonts}
\usepackage{siunitx}

\usepackage{etoolbox}

\newcommand{\um}{\,\textmu\text{m}\xspace}

\newcommand{\ke}{\,\text{ke}\ensuremath{^-}\xspace}
\newcommand{\aunit}[1]{\ensuremath{\text{\,#1}}}
\newcommand{\mevc}{\ensuremath{\aunit{Me\kern -0.1em V\!/}c}\xspace}
\newcommand{\gevc}{\ensuremath{\aunit{Ge\kern -0.1em V\!/}}c\xspace}
\newcommand{\mevcc}{\ensuremath{\aunit{Me\kern -0.1em V\!/}c^2}\xspace}
\newcommand{\gevcc}{\ensuremath{\aunit{Ge\kern -0.1em V\!/}c^2}\xspace}

\glsdisablehyper
\newacronym{asic}{ASIC}{application specific integrated circuit}
\newacronym{hl-lhc}{HL-LHC}{High Luminosity Large Hadron Collider}
\newacronym{velo}{VELO}{Vertex Locator}

\newacronym{drie}{DRIE}{deep reactive ion etching}
\newacronym{3d-ddtc}{3D-DDTC}{3D double-side double type column}
\newacronym{dut}{DUT}{detector under test}
\newacronym{mcp-pmts}{MCP-PMTs}{microchannel plate photomultipliers}
\newacronym{vco}{VCO}{voltage controlled oscillator}
\newacronym{toa}{ToA}{Time of Arrival}
\newacronym{ftoa}{fToA}{fine ToA}
\newacronym{uftoa}{ufToA}{ultra fine ToA}
\newacronym{tot}{ToT}{Time over Threshold}
\newacronym{tdc}{TDC}{time to digital converter}
\newacronym{mpv}{MPV}{most probable value}

\newacronym{4d}{4D}{four-dimensional}
\newacronym{1e}{1E}{one electrode}
\newacronym{2e}{2E}{two electrode}

\title{Testbeam characterization of a 3D silicon sensor read out by Timepix4}

\makeatletter
\renewcommand{\thefootnote}{\fnsymbol{footnote}}
\patchcmd{\maketitle}
  {\bfseries\raggedright\sffamily\the\auth@toks}
  {\bfseries\raggedright\sffamily\small\the\auth@toks}
  {}{}
\makeatother

\author[a,*]{E.~Chatzianagnostou\note{Corresponding author}}
\author[b]{M.~J.~Madurai}

\author[a]{K.~Akiba}
\author[c]{D.~Bacher}
\author[d]{R.~Bates}
\author[a]{M.~van~Beuzekom}
\author[a]{T.~Bischoff}
\author[e]{V.~Coco}
\author[e]{R.~Dumps}
\author[a]{T.~Evans}
\author[a]{K.~Heijhoff}
\author[f]{D.~Johnson}
\author[a]{U.~Krämer}
\author[e]{E.~Lemos~Cid}
\author[a]{D.~Oppenhuis}
\author[e]{T.~Pajero}
\author[e,g]{E.~Rodríguez~Rodríguez}
\author[h]{D.~Rolf}

\affiliation[a]{{Nikhef, Science Park 105, 1098 XG Amsterdam, the Netherlands}}
\affiliation[b]{{Department of Physics, University of Warwick, Coventry, United Kingdom}}
\affiliation[c]{{Department of Physics, University of Oxford, Denys Wilkinson Bldg., Keble Road, Oxford, OX1~3RH, United Kingdom}}
\affiliation[d]{{University of Glasgow, Department of Physics and Astronomy, Glasgow, G12 8QQ, Great Britain}}
\affiliation[e]{{CERN, Esplanade des Particules 1, 1211 Geneva, Switzerland}}
\affiliation[f]{{School of Physics and Astronomy, University of Birmingham, Edgbaston, Birmingham, B15 2TT, United Kingdom}}
\affiliation[g]{{Instituto Galego de Fisica de Altas Enerxias (IGFAE), Universidade de Santiago de Compostela, Santiago de Compostela, Spain}}
\affiliation[h]{{TU Dortmund, Otto-Hahn-Strasse 4, 44227 Dortmund, Germany}}

\emailAdd{echatzia@nikhef.nl}

\abstract{
 Testbeam results from a 300\um-thick 3D silicon sensor bump-bonded to a Timepix4 ASIC, are presented. The hit detection efficiency, spatial resolution, and timing performance of the 3D sensor are studied for several track angles, bias voltages and charge thresholds. The time measurements are corrected for Timepix4 clock-frequency variations and timewalk, while for the spatial studies, nonlinear charge-sharing corrections are determined. At perpendicular incidence, the time resolution of the 3D detector is equal to 245\,ps with a $\SI{97}{\percent}$ hit detection efficiency. An optimal angle of $\ang{8}$ with respect to the beam direction of the 3D detector was found, in which the time resolution improves by $\SI{6}{\percent}$ compared to normal incidence, while the hit detection efficiency reaches above $\SI{99}{\percent}$ and the spatial resolution is approximately 7\um. Intrapixel studies show that a time resolution of 153\,ps at the most probable value of the signal charge can be achieved for hits  between the electrodes at normal incidence, while the timing performance of these best-performing regions deteriorate upon sensor rotation. The timing properties at different depths of the 3D sensor have been investigated with tracks at grazing-angle incidence, revealing a dependence of the time resolution along the sensor depth.
}


\arxivnumber{}

\compress

\begin{document}
\maketitle
\flushbottom


\renewcommand{\thefootnote}{\arabic{footnote}}

\section{Introduction} \label{sec:intro}

The upcoming \gls{hl-lhc}~\cite{HL-LHC} will pose significant challenges to particle detectors, which will have to cope with the increased number of quasisimultaneous proton-proton collisions. In particular, the LHCb Upgrade~II \gls{velo}~\cite{VELO_U2} will require excellent spatial and temporal resolutions to correctly associate tracks with the collision vertices where they were produced. A per-hit time measurement with a resolution on the order of 50\,ps~\cite{LHCbVELOgroup:2022vrz}, enabling \gls{4d} tracking, and a spatial resolution better than 10\um are targeted. Both are essential to discriminate between primary and displaced secondary vertices in such an environment, which are a distinct feature of beauty and charm hadron decays. These requirements, together with the high radiation levels expected at the \gls{hl-lhc}, motivate the exploration of fast and radiation-hard sensor technologies. Among the proposed solutions under consideration for the Upgrade~II \gls{velo}, three-dimensional (3D) silicon sensors constitute a promising sensor technology for fast and radiation-tolerant tracking detectors.

Silicon sensors with 3D electrode configuration have electrodes vertically embedded into the silicon bulk. In this way, the drift is decoupled from the substrate thickness, allowing for shorter drift lengths of the charge carriers without compromising the signal amplitude. In 3D sensors, the pixel pitch determines the drift distance, which can be minimized to ensure high charge collection efficiency even in the presence of severe charge trapping in the bulk. Three-dimensional sensors have demonstrated radiation hardness up to fluences of 1$\times$10$^{17}$~1~MeV~n\textsubscript{eq}~cm$^{-2}$~\cite{radiation_hard_3d}. 

Three-dimensional pixel sensors are inherently fast devices, since the short drift path of charge carriers, perpendicular to the electrodes, also results in fast charge collection times. By decoupling the directions of charge deposition and collection, Landau fluctuations in the total generated charge take place along the direction of the track and to first order only affect the signal amplitude, which can be accounted for by timewalk corrections. Therefore, the impact of Landau fluctuations on timing performance, also known as Landau noise~\cite{riegler}, is less significant in 3D silicon sensors. An important limitation of 3D sensors with columnar electrodes arises from spatial non-uniformities in the weighting field~\cite{ramo}. In this respect, more uniform weighting fields can be achieved by replacing columnar electrodes with trenches or by exploring multiple readout 3D electrode configurations, such as the \gls{2e} geometry~\cite{1e2e}.

Single-pixel 3D test structures with fast discrete readout channels have demonstrated time resolutions on the order of tens of picoseconds~\cite{3d_time_resol, radiation_hard_3d}, while the scalability and maintenance of this performance in pixelated devices integrated with a readout chip remain to be studied. In 3D sensors, the length and diameter of the electrodes, which are correlated by the achievable aspect ratio~\cite{drie}, increase the sensor capacitance and consequently the electronics jitter.  
Optimal design and development of a 3D sensor with enhanced timing capabilities and minimal capacitance are therefore of great importance, and efforts are being made in this direction~\cite{tcode_simul,3d_simul_old}.

This article presents the spatial and timing performance of a 3D silicon sensor~\cite{3D_Glasgow} bump bonded to, and read out by, a Timepix4 \gls{asic}~\cite{tpx4}. The characteristics of the specific 3D sensor and the key features of the Timepix4 \gls{asic} are introduced in \cref{sec:dut}, while in \cref{sec:setup} the experimental setup is described. In \cref{sec:spatial} the application of $\eta$~corrections and the resulting spatial resolution of the 3D sensor are reported. A comprehensive  study of the timing performance is presented in \cref{sec:time}, where the time resolution for particles crossing at perpendicular, small-angle, and grazing-angle incidences are discussed in detail. The conclusions are summarized in \cref{sec:conclusions}.

\section{Detector under test} \label{sec:dut}

\begin{figure}
    \centering
    \includegraphics[width=0.4\textwidth]{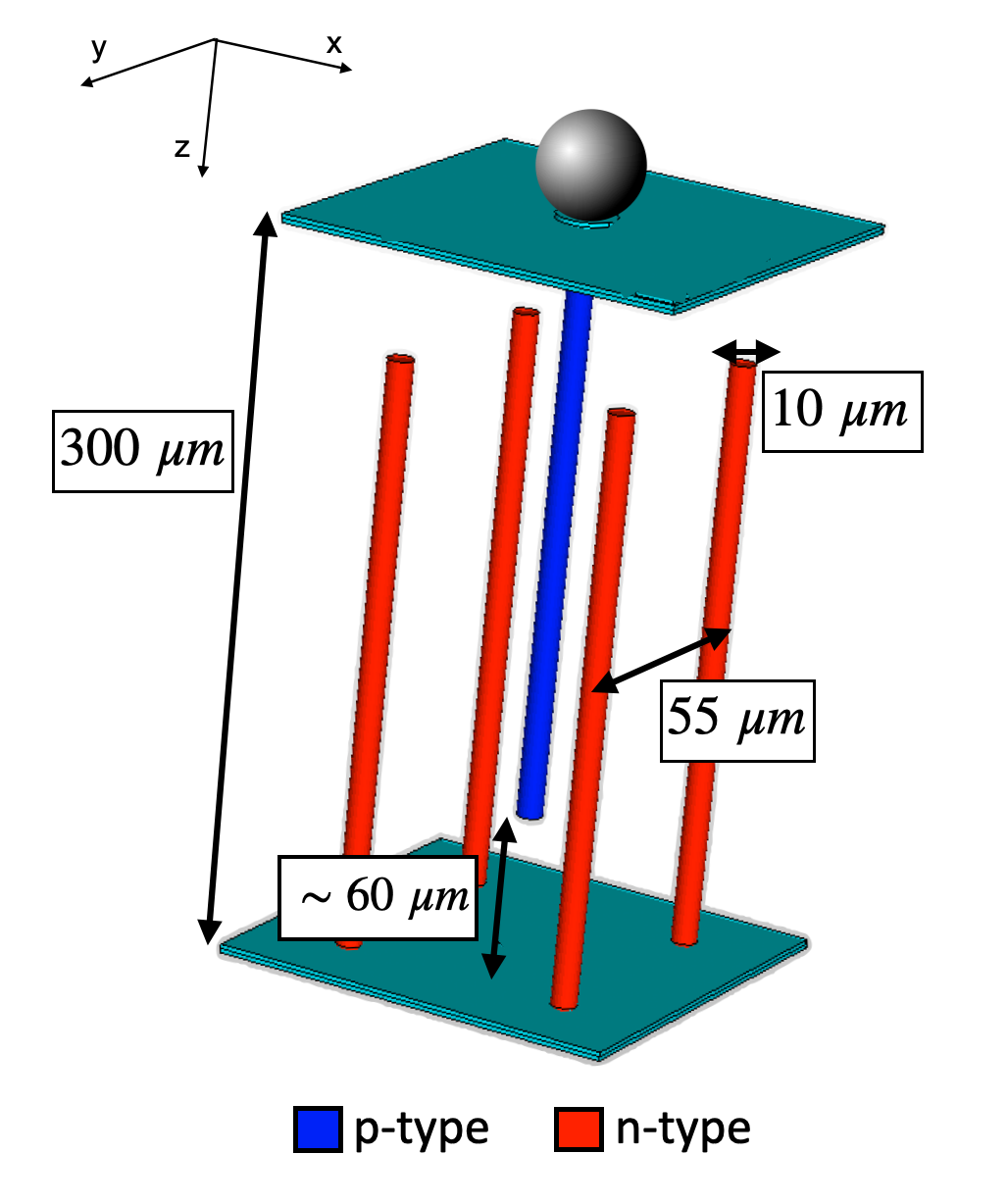}
    \caption{A schematic illustration of the \glsentryshort{3d-ddtc} silicon sensor under test.  \label{fig:3D_scheme}}
\end{figure}

The fabrication of 3D silicon sensors is based on \gls{drie}~\cite{drie}, where cylindrical holes or trenches are etched through the wafer and filled with heavily doped p-type and n-type material to form an active edge electrode. The 3D sensor used in this study is a \gls{3d-ddtc} sensor with partially through electrodes~\cite{3d-ddtc} and \gls{1e} configuration~\cite{1e2e} fabricated at IMB-CNM~\cite{3DatCNM}. A schematic illustration of the sensor is presented in \cref{fig:3D_scheme}. From the front side, the etched holes are filled with highly p-doped polysilicon and, since they are formed on a 300\um Si-n substrate, they are referred to as junction electrodes. The highly n-doped pillars from the backside are called ohmic electrodes. An active silicon depth of 70\um below the junction electrodes and 60\um below the ohmic electrodes is present. With a nominal diameter of the etched hole of 10\um, a high aspect ratio was achieved, minimizing the ineffective regions within the sensor volume. The ohmic pillars on the backside of the sensor are interconnected through a highly n-doped polysilicon layer, forming a common electrode to which the bias voltage is applied. The frontside electrodes form pixels that are individually connected to the readout channels of the Timepix4 \gls{asic}~\cite{tpx4}. For this reason, the junction electrodes are hereafter referred to as readout electrodes and the ohmic ones as field electrodes. The 3D sensor under study was initially bonded to a Timepix3 \gls{asic} and the results of its performance can be found in~\cite{3DonTPX3}. Recently, it was transplanted to a Timepix4 \gls{asic} which is read out by a SPIDR4 readout system~\cite{spidr}.

The Timepix4 \gls{asic} is composed of $512\times448$ square pixels with 55\um pitch. For this study, the Timepix4 is operated in the ToA \& ToT mode, in which both the \gls{toa} and \gls{tot} are registered for each hit. The determination of time is based on a system of local \glspl{vco} shared by groups of $4\times2$ adjacent pixels, known as superpixels, running at a nominal frequency of 640\,MHz.
The 640\,MHz \gls{vco} is activated when the preamplifier output crosses a set threshold value, and oscillates until the next rising edge of the 40\,MHz global clock. The number of 640\,MHz clock cycles is counted, providing a \gls{ftoa} measurement with a nominal \gls{tdc} bin size of approximately 1.56\,ns. To improve the time resolution, four phase-shifted copies of the \gls{vco} are generated and their phases are captured at the rising edge of the 40\,MHz global clock. In this way, the 1.56\,ns period is further divided into eight bins, resulting in an \gls{uftoa} measurement with a granularity of 195\,ps. Therefore, Timepix4 provides a nominal \gls{tdc} resolution of 56.4\,ps, with $60.3\pm1.0$\,ps being the best achieved~\cite{tpx4_resol}.

For each hit, \gls{tot} is measured by counting the number of the 40\,MHz clock cycles in which the signal remains above the threshold. The \gls{vco} is activated once again when the preamplifier output drops below the threshold, allowing a \gls{tot} measurement precision of 1.56\,ns. Due to the nonlinear response of the front-end electronics in the energy range close to the threshold, the relationship between the registered \gls{tot} value and the charge $q$ generated in the pixel by an incoming particle is described by a surrogate function~\cite{tot_to_q_eq}, given as the inversion of \cref{eq:tot-charge}.
\begin{equation}
\label{eq:tot-charge}
\mathrm{ToT} = p_0 + p_1\cdot q - \frac{p_2}{q-p_3}\,.
\end{equation}
The parameters of \cref{eq:tot-charge} are determined independently for each pixel through a charge calibration procedure which is performed by controlled charge injection (test pulses) into the analog front-end of each pixel in the range [$1\ke$, $16\ke$], with a step of $0.2\ke$.

\section{Testbeam setup} \label{sec:setup}

The present analysis is based on data collected during August 2024 and May 2025 at the SPS H8 beam-line facility at CERN~\cite{sps-h8}. The \gls{dut} was exposed to a mixed-hadron ($p,\ \pi,\ K$) beam at 180\gevc, delivered in 4.5 second-long spills containing a few million particles each. The Timepix4 beam telescope~\cite{telescope} was used as the track reconstruction system, providing a $2.3\pm0.1$\um pointing resolution at the \gls{dut} position. A composition of stages at the center of the telescope, which enable $xy$ translation and rotation around $y$-axis (angle $\phi$) and  $x$-axis (angle $\theta$), hosts the \gls{dut}. A time-reference system composed of two \gls{mcp-pmts}~\cite{mcp-pmt} achieves a timestamp accuracy of about 12\,ps, and is used to characterize the temporal performance of the \gls{dut}. The reconstruction of the telescope tracks is performed by the Kepler package, described in~\cite{kepler, telescope}.

\section{Spatial resolution} \label{sec:spatial}

The spatial resolution depends on a variety of factors related to the design of the 3D sensor, namely the pixel pitch and sensor thickness, but also on operational and external parameters, such as the charge threshold of the Timepix4 \gls{asic}, the bias voltage applied to the sensor and the incident angle of the impinging charged-particle track. Understanding the spatial resolution of the sensor as a function of these parameters provides essential context for the timing analysis that follows.

For a charged particle traversing a silicon sensor, the generated electron-hole (eh) pairs may be collected by multiple neighboring pixels, which are grouped together following the algorithm described in~\cite{telescope} to form a cluster.
The spatial resolution of the 3D sensor is estimated from the spatial residuals, defined as the difference between the intercept of the reconstructed incident track with the sensor and the charge-weighted position of the associated cluster. The width of this distribution is a convolution of the intrinsic spatial resolution of the 3D sensor and the pointing resolution of the telescope. Hereafter, the quoted spatial resolution is calculated as the standard deviation of the central $99\%$ of the residual distribution, with the pointing resolution of the telescope subtracted in quadrature.

\begin{figure}
    \centering
    \includegraphics[width=0.49\linewidth]{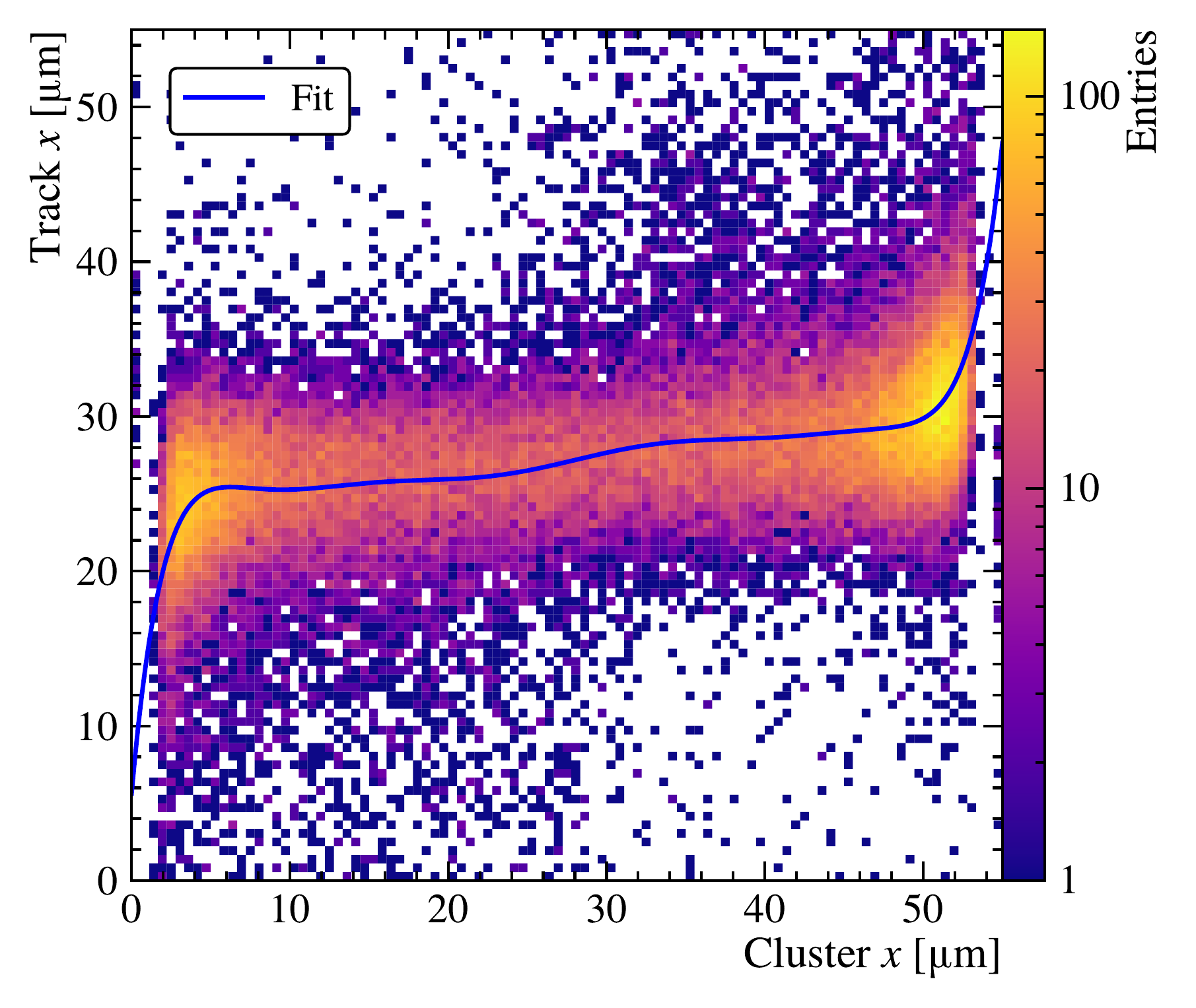}
    \includegraphics[width=0.49\linewidth]{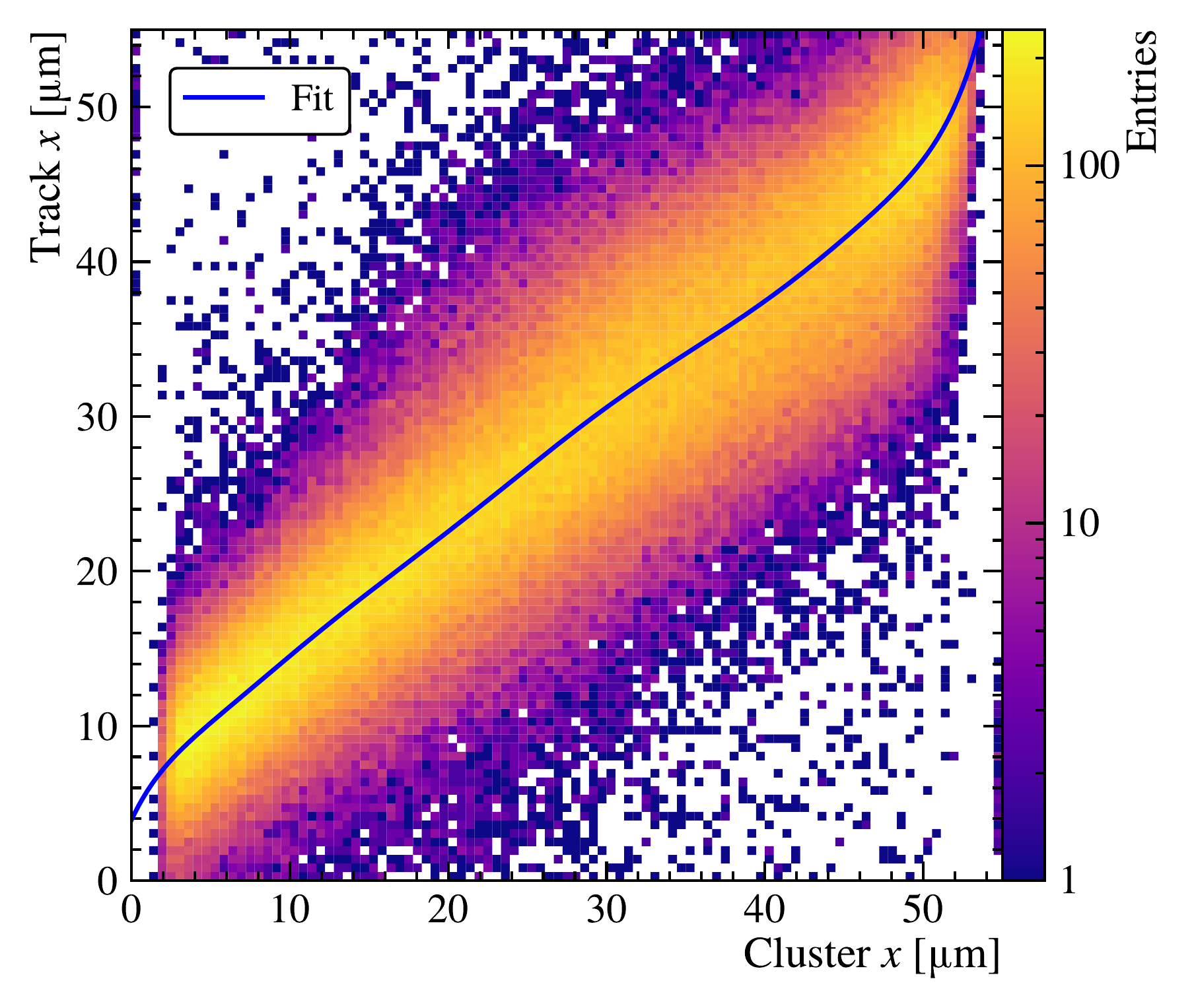}
    \caption{The reconstructed track position compared to the charge-weighted cluster position, before the $\eta$ correction, for two-pixel clusters from a perpendicularly incident charged-particle track (left) and a track inclined by \ang{8} (right). The polynomial fit used to derive the $\eta$ correction is overlaid in blue. The boundary between two adjacent pixels corresponds to 27.5\um.}
    \label{fig:charge_sharing_corrections}
\end{figure}

\begin{figure}
    \centering
    \includegraphics[width=0.49\linewidth]{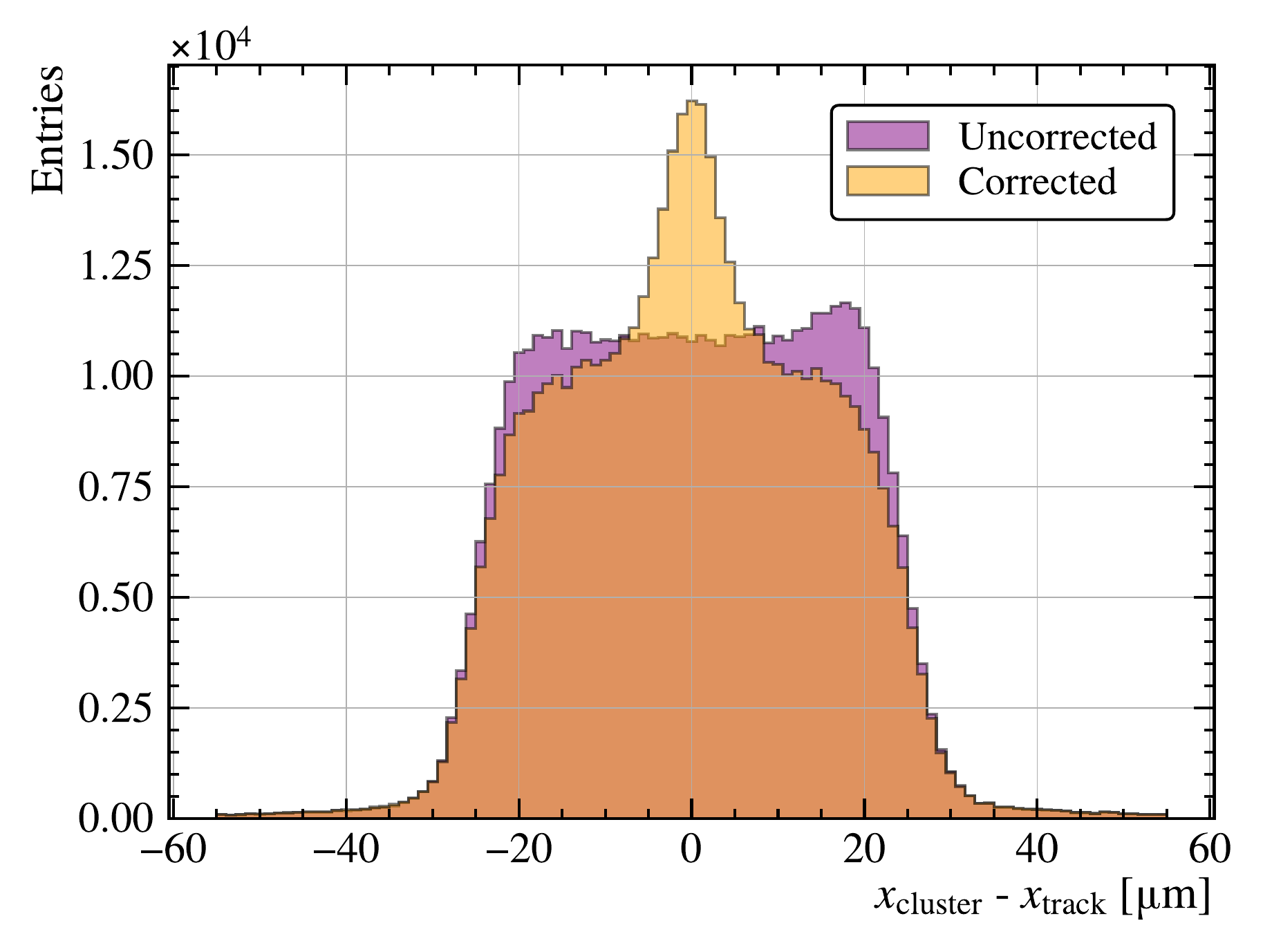}
    \includegraphics[width=0.49\linewidth]{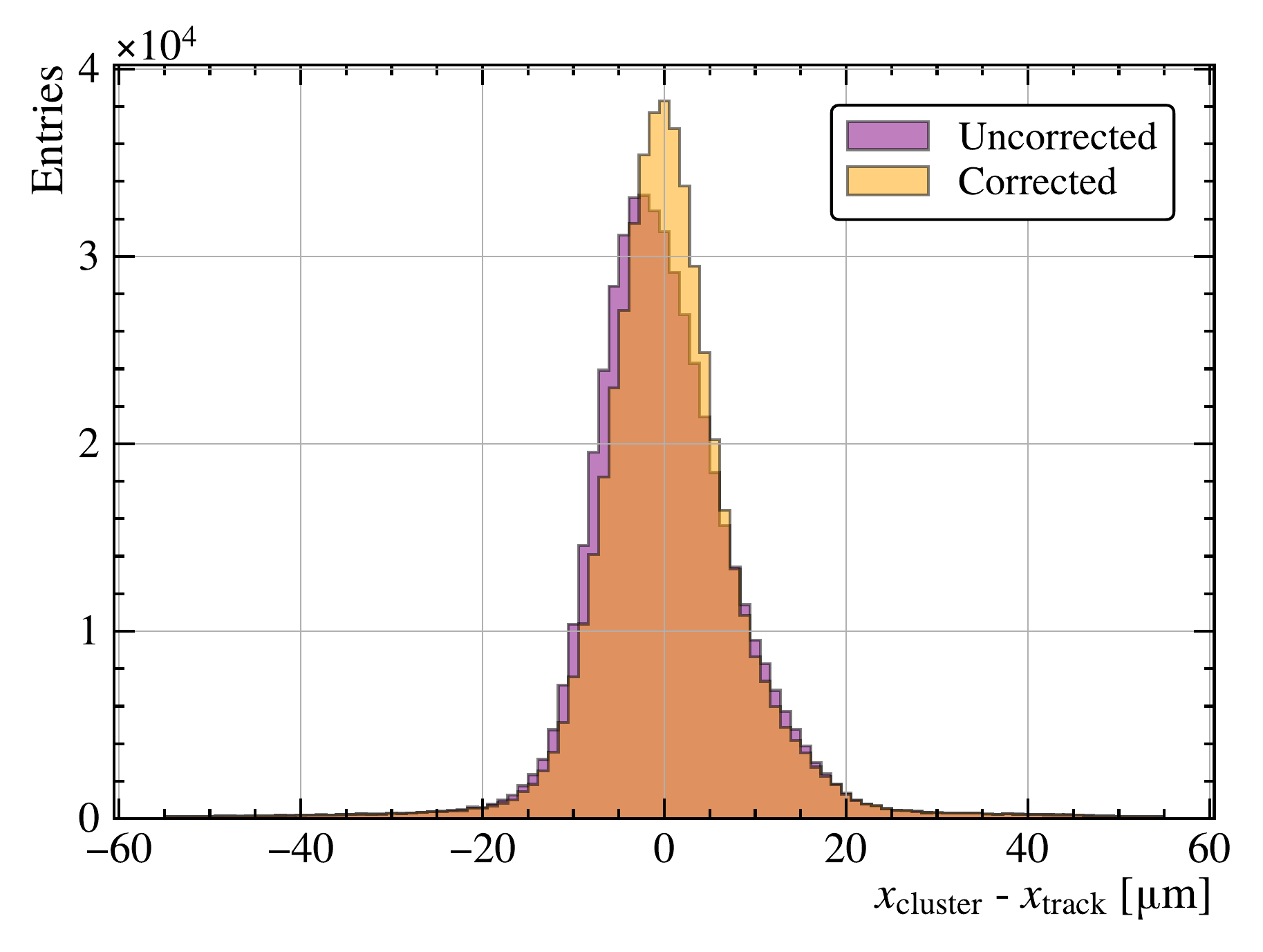}
    \caption{Distribution of spatial residuals for clusters up to a maximum size of two in the $x$-direction before (blue) and after correction (orange) for perpendicularly incident tracks (left) and tracks inclined by \ang{8} (right).}
    \label{fig:x_residual_corrections}
\end{figure}

The charge-weighted average of the hits in a cluster is a biased estimator of the true position at which a charged-particle track traversed the sensor~\cite{Turchetta:1993vu}. This bias arises from inhomogeneities in the weighting field and the random nature of the charge-carrier diffusion, which results in a nonlinear relationship between the deposited charge and the measured pixel response. At small incident angles the induced signal in subleading pixels is predominantly from the lateral diffusion of the charge relative to the field, meaning that the charge sharing is particularly sensitive to these nonlinearities and results in a larger bias. Conversely, at larger angles a greater fraction of the signal in subleading\footnote{The pixel that registers the largest amount of charge in a given cluster is referred to as the leading pixel, while all remaining pixels are referred to as subleading pixels.} pixels is from charge liberated directly within those pixels, resulting in a smaller bias and a more linear relationship. \Cref{fig:charge_sharing_corrections} compares the track position to the charge-weighted position for associated two-pixel clusters for both perpendicular and angled incident charged-particle tracks. Following the method in~\cite{Akiba:2011vn}, a polynomial fit is performed to the average charge-weighted position as a function of intrapixel position, where the order of the polynomial is selected as that which minimizes the sum of squared residuals per degree of freedom of the fit, up to a maximum degree of nine. The resulting fit is used to correct for the observed nonlinearities, a procedure known as $\eta$ corrections, and is repeated separately for all clusters containing up to two hits in the $x$ and $y$ directions at each incident angle of the impinging charged-particle track. \Cref{fig:x_residual_corrections} shows the distributions of spatial residuals before and after the corrections calculated from the fits displayed in \cref{fig:charge_sharing_corrections}, where a reduction in standard deviation of 7\% and 12\% can be seen for charged particles of perpendicular and angled incidence, respectively. The top-hat distribution in the residuals at perpendicular incidence is a result of the binary resolution of single-pixel clusters, while the central peak appears after the $\eta$ correction process to two-pixel clusters in which there is a large charge imbalance between the leading and subleading pixels. The asymmetry observed in the residuals distribution at angled incidence is a result of residual nonlinearities that are not fully corrected in the $\eta$ correction process.

\begin{figure}
    \centering
    \includegraphics[width=0.5\linewidth]{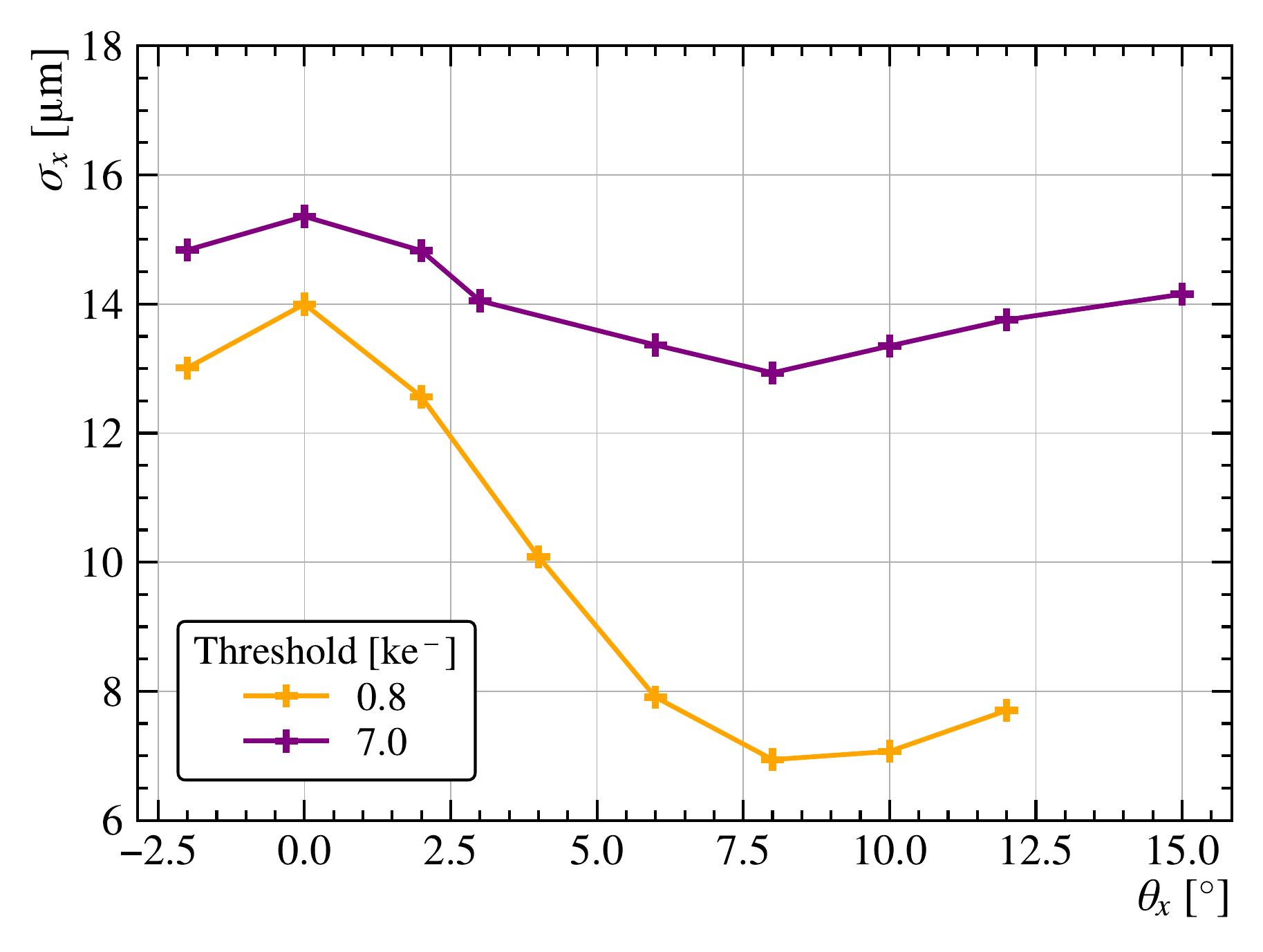}
    \caption{Spatial resolution along the $x$-direction of the 3D sensor as a function of charged-track incident angle relative to the $x$-axis for two values of threshold. A reverse bias of 60\,V was applied to the 3D sensor.}
    \label{fig:spin_scan}
\end{figure}

%
The spatial resolution along the $x$-direction as a function of track incidence angle for two values of threshold is shown in \cref{fig:spin_scan}. The lower threshold of $0.8\ke$ yields a significantly better resolution across all angles, as charge shared between neighbouring pixels that would otherwise fall below threshold remains detectable, allowing the charge-weighted clustering algorithm and $\eta$ correction process to more accurately reconstruct the track position. Furthermore, at higher thresholds the spatial resolution of single-pixel clusters is degraded due to an increased number of hits falling below threshold within the region below the readout electrode, resulting in double-peaked spatial residual distribution. At perpendicular incidence, the resolution is approximately $14$\um and $15$\um for the low and high thresholds, respectively.
For the minimum threshold value with workable noise levels for this sensor ($0.8\ke$) the spatial resolution improves with increasing spin angle, reaching a minimum of approximately $7$\um at a track incidence angle of around \ang{8}, where charge sharing between two pixels is maximized. Beyond this angle the spatial resolution degrades slightly. In contrast, the spatial resolution at the high threshold of $7.0\ke$ is much less sensitive to incident track angle, varying only between approximately $13$\um and $15$\um. This is again due to the charge shared in the subleading pixels being below threshold, reducing most clusters to single-pixel hits for which the charge-weighting clustering algorithm cannot be used, limiting the resolution to the binary single-pixel contribution.

\begin{figure}
    \centering
    \includegraphics[width=0.5\linewidth]{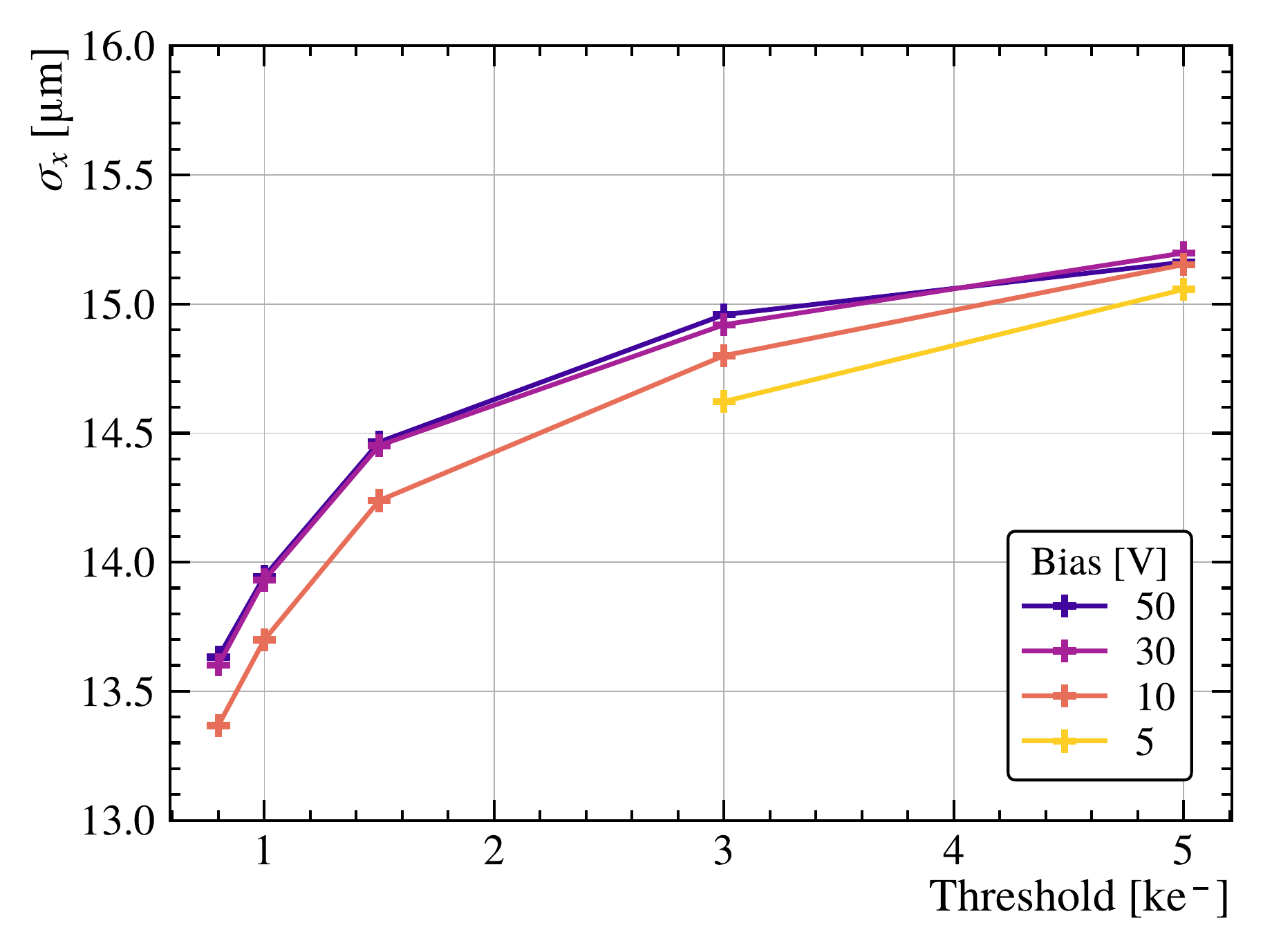}
    \caption{Spatial resolution along the $x$-direction at varying values of threshold and reverse bias for perpendicularly incident tracks.}
    \label{fig:threshold_bias_scan}
\end{figure}

%
The spatial resolution along the $x$-direction at varying values of threshold and reverse bias for perpendicularly incident tracks is shown in \cref{fig:threshold_bias_scan}. Due to the short inter-electrode distance in the 3D sensor, the depletion voltage is very low, and the sensor is fully depleted from about 5\,V onwards. Consequently, the charge sharing and charge collection properties of the 3D sensor are mostly saturated beyond a bias voltage of 40\,V, leaving the threshold as the dominant factor affecting the spatial resolution. The spatial resolution degrades with increasing threshold, from around $13.5$\um at a threshold of $0.8\ke$ to approximately $15$\um at $5\ke$, again as a result of the progressive reduction of charge-sharing information as the shared charge in subleading pixels increasingly falls below threshold. Altering the reverse bias applied to the 3D sensor has a small effect on the spatial resolution at low threshold values, with this effect becoming increasingly smaller at higher thresholds. At low threshold, this is as a result of charge sharing in the neighbouring pixels through the lateral diffusion of the charge carriers as they drift towards the electrodes. The amount of lateral diffusion depends on the drift time of the charge carriers, which up to saturation, is governed by the electric field strength. This in turn, once the sensor is fully depleted, is proportional to the bias voltage applied. At higher bias voltages the drift velocity is greater, charge carriers reach the electrodes faster, and the amount of lateral diffusion is reduced. This results in a slight reduction in charge sharing and therefore a degraded spatial resolution at low threshold. This is true up to saturation of the charge carrier velocity, which occurs at approximately 40\,V in this sensor, where increasing the bias voltage further has no effect on the spatial resolution. At higher thresholds, this effect is not observed, as the shared charge in subleading pixels is below threshold, regardless of the amount of lateral diffusion. Overall, the impact of the applied bias voltage is small.

\section{Time resolution} \label{sec:time}

The time resolution, $\sigma_t$, is calculated from the RMS of the central $\SI{95}{\percent}$ of the time residual distribution, as will be discussed in \cref{subsec:perpendicular}. The time residual is defined as the difference between the measured hit time of the leading pixel of the \gls{dut} cluster and the reference time, $t_\text{ref}$. The \gls{dut} time measurement is corrected for \gls{vco} frequency variations and timewalk. The reference time is calculated as the average timestamp of the two \gls{mcp-pmts} with an uncertainty of approximately 12\,ps that is neglected hereafter.

\begin{figure}
    \centering
    \includegraphics[width=.42\textwidth]{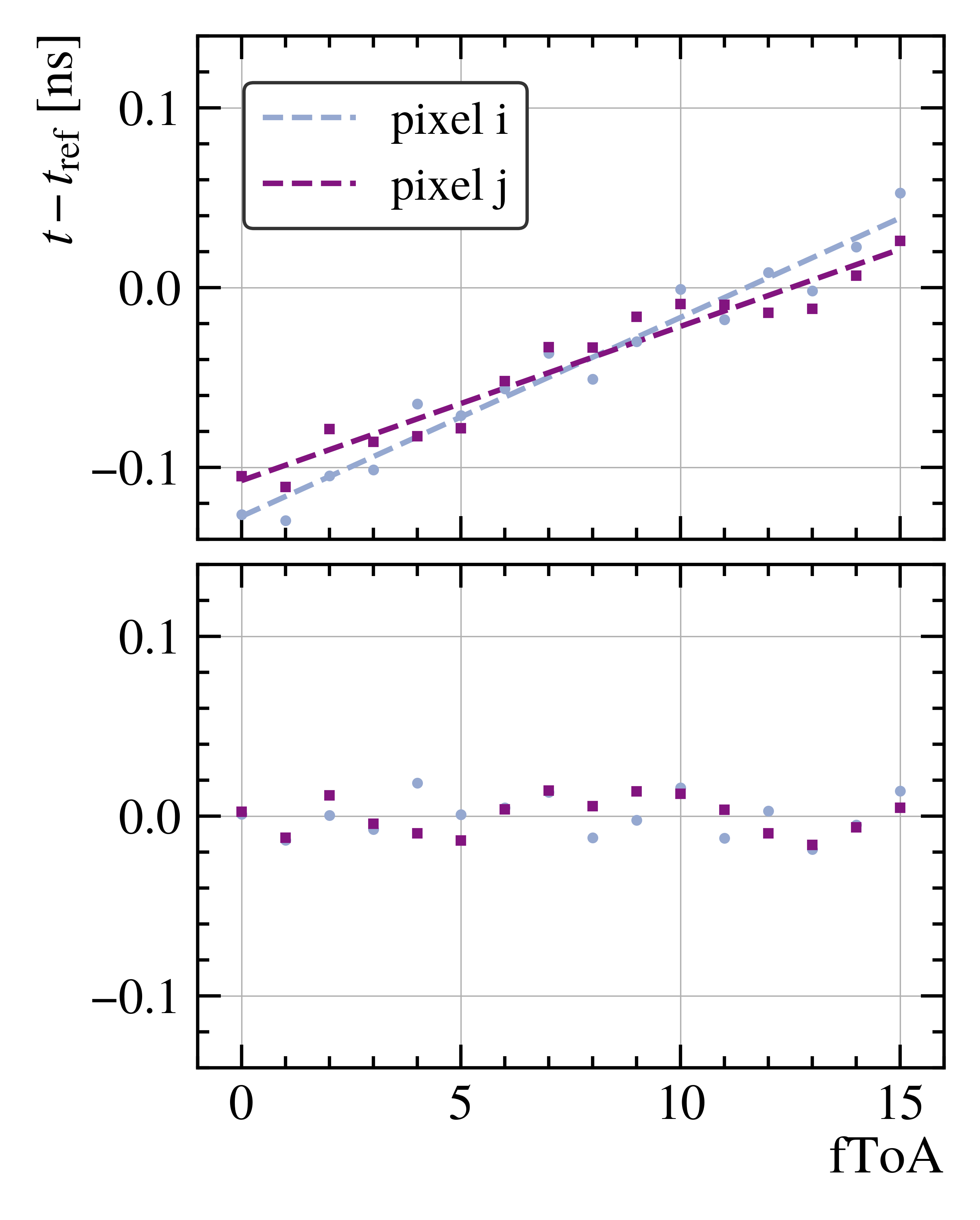}
    \caption{Average time difference between the measured and the reference time for two example pixels (top) before and (bottom) after \gls{vco} corrections.  \label{fig:vco_corr}}
\end{figure}

A 640\,MHz \gls{vco} divides the 25\,ns period of the 40\,MHz clock into 16 \gls{tdc} bins providing the \gls{ftoa} measurement, with a granularity of 1.56\,ns, as mentioned in \cref{sec:dut}. In an ideal Timepix4 \gls{asic}, all \gls{vco} copies in the superpixels would be identical. However, variations in the process parameters of the \glspl{vco} during the fabrication of Timepix4, as well as voltage and temperature differences across the chip, lead to deviations from the nominal frequency. For this \gls{dut}, an average frequency of $634$\,MHz is measured with $7$\,MHz spread across the detector area. This frequency spread generates differences in the time measurements of up to 0.5\,ns. The \gls{vco} frequency deviations can be calculated using testbeam data for pixels with a sufficient number of hits. In \cref{fig:vco_corr}, the time difference between the measured hit time and the reference time in two example pixels is shown as a function of the \gls{ftoa} value. Assuming a superpixel frequency of 640\,MHz, all points would have the same average time residual, within the statistical error, which would be associated with the pixel time offset. If the \gls{vco} runs slightly faster (slower) than 640\,MHz, the \gls{tdc} bins shrink (expand), more (fewer) oscillating cycles are completed, and the hit is falsely reported at an earlier (later) time. The relationship between the relative time delay and the \gls{ftoa} values can be parametrized as
\begin{equation}
\label{eq:vco_corr}
\Delta t_\text{vco} = t -t_\text{ref} = \alpha\frac{1.56\ \text{ns}}{16}\text{fToA} +\beta\,,
\end{equation}
where $\alpha$ and $\beta$ are parameters fit to data (see \cref{fig:vco_corr}). The advantage of determining \gls{vco} corrections for each pixel instead of superpixel is that second-order differences, such as pixel offsets (parameter $\beta$), can be determined. However, this approach requires sufficiently large samples of calibration data per pixel.

After the application of \gls{vco} corrections, time residuals must also be corrected for the timewalk\footnote{The term timewalk refers to the charge dependence of the measured hit time introduced by the constant threshold discriminator. However, different effects such as variations in the drift velocity of the charge carriers, may contribute to the measured timewalk and cannot be resolved in this analysis.} effect in order to determine the optimal time resolution of the 3D detector. The relationship between charge, $q$, and the observed difference between the measured time, $t$, and $t_\text{ref}$ is modeled by the function
\begin{equation}
\label{eq:tw_corr}
\Delta t_\text{tw} = t -t_\text{ref} - \Delta t_{\text{vco}} = \frac{a}{(q+b)^c} +d\,,
\end{equation}
where the empirical parameters $a$, $b$, $c$ and $d$ are fit to data.
Typically, the timewalk correction is modeled with a single set of parameters for the entire detector (hereafter global timewalk correction) $\Delta t_\text{tw}(q)$ or for a separate set of parameters for every pixel (per-pixel timewalk correction) $\Delta t_\text{tw}(q,\text{row},\text{col})$. However, in the following analysis, a different approach for the timewalk correction, based on the hit position within the pixel (position-informed timewalk correction) $\Delta t_\text{tw}(q,x,y)$, will be introduced and its performance will be discussed.

\subsection{Perpendicular incidence}\label{subsec:perpendicular}

\begin{figure} 
    \centering
    \includegraphics[width=.5\textwidth]{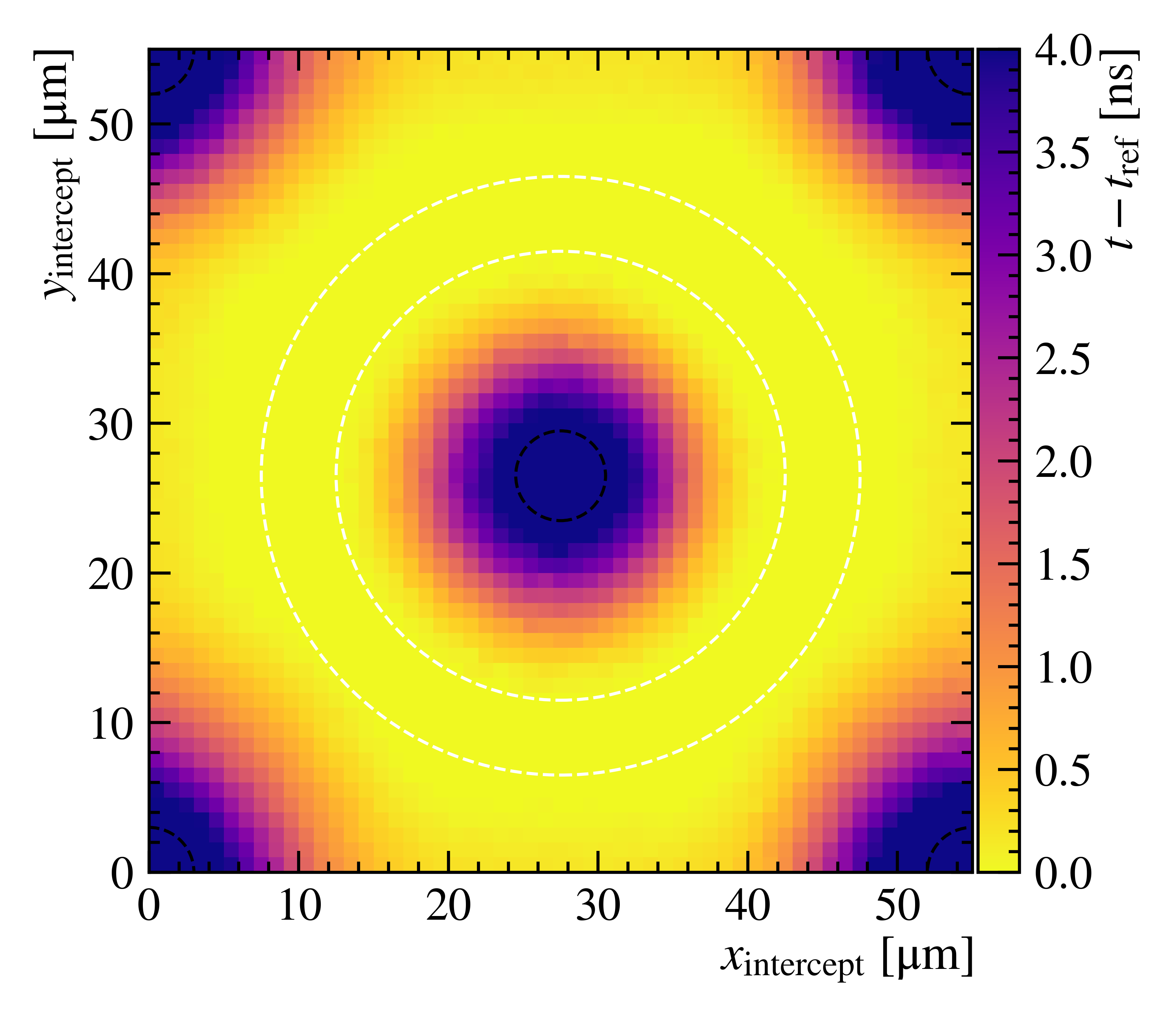}
    \caption{Relative time delay after \gls{vco} corrections. 
    \label{fig:intrapixel}}
\end{figure}

For incident charged particles, perpendicular incidence is determined by the set of ($\theta_0,\phi_0$) angles of the rotation stages for which the mean cluster size is minimized. The mean cluster size at normal incidence is equal to $1.23$ pixels, indicating that the geometry of the 3D sensor does not favor charge sharing. Since a significant percentage of a given cluster's charge is captured by a single pixel, the pixel with the first registered time measurement in the cluster (also leading pixel for $>\SI{99}{\percent}$ of the cases) is used to determine the time resolution.

The average relative delay between the \gls{dut} hit time, corrected for the \gls{vco} variations, and the reference time is shown as a function of the incident track position within the pixel in \cref{fig:intrapixel}. The readout and field electrodes are identified as the regions within the pixel that exhibit large positive time delay.
The large time delay below the electrodes is associated with the low charge generated in the corresponding regions, which is accounted for by the timewalk correction. In addition to this effect, the contribution of depth variations in the electric field to the drift time is discussed in \cref{subsec:grazing}.
Regions of the pixel cell with and without electrodes are drawn as contour lines in \cref{fig:intrapixel} and are defined conservatively to minimize contributions from transition regions and isolate the characteristic behavior of the regions of interest.

The \gls{mpv} of the hit charge distribution depends on the particle's impact position within the pixel. The \gls{mpv} is measured to be $19.3\ke$ for events in the region between the electrodes, and $4.0\ke$ and $4.6\ke$ in the region of field electrodes and readout electrodes, respectively. These values are around $\SI{15}{\percent}$ lower than the expected charge deposition of minimum ionizing particles at normal incidence in a 300\um thick silicon sensor. 
The deviation between the measured and expected \glspl{mpv} can be explained either by an active thickness of the 3D sensor less than 300\um or by a bias in the charge calibration. The active thickness of the sensor is investigated by studying the cluster length of particles passing through the sensor at grazing angles in the range [\ang{80},\,\ang{89}]~\cite{grazing_method} and is measured to be $301.9\pm3.4$\um, compatible with the nominal sensor specifications. A mismatch in charge calibration can be quantified by analyzing the spectrum of a {\textsuperscript{241}Am} radioactive source. The primary $\gamma$-ray peak at 59.54\,keV~\cite{NNDC_Am241} is translated into $16.5\ke$, assuming an average ionization energy of 3.6\,eV in silicon. However, in the \gls{dut} this peak appears to be at $13.6\ke$, revealing a $\SI{17.5}{\percent}$ difference. This discrepancy can be attributed, at least partially, to the tolerance of the calibration capacitors in Timepix4, but a more thorough investigation is required to understand this effect.

\begin{figure}
    \centering
    \includegraphics[width=.5\textwidth]{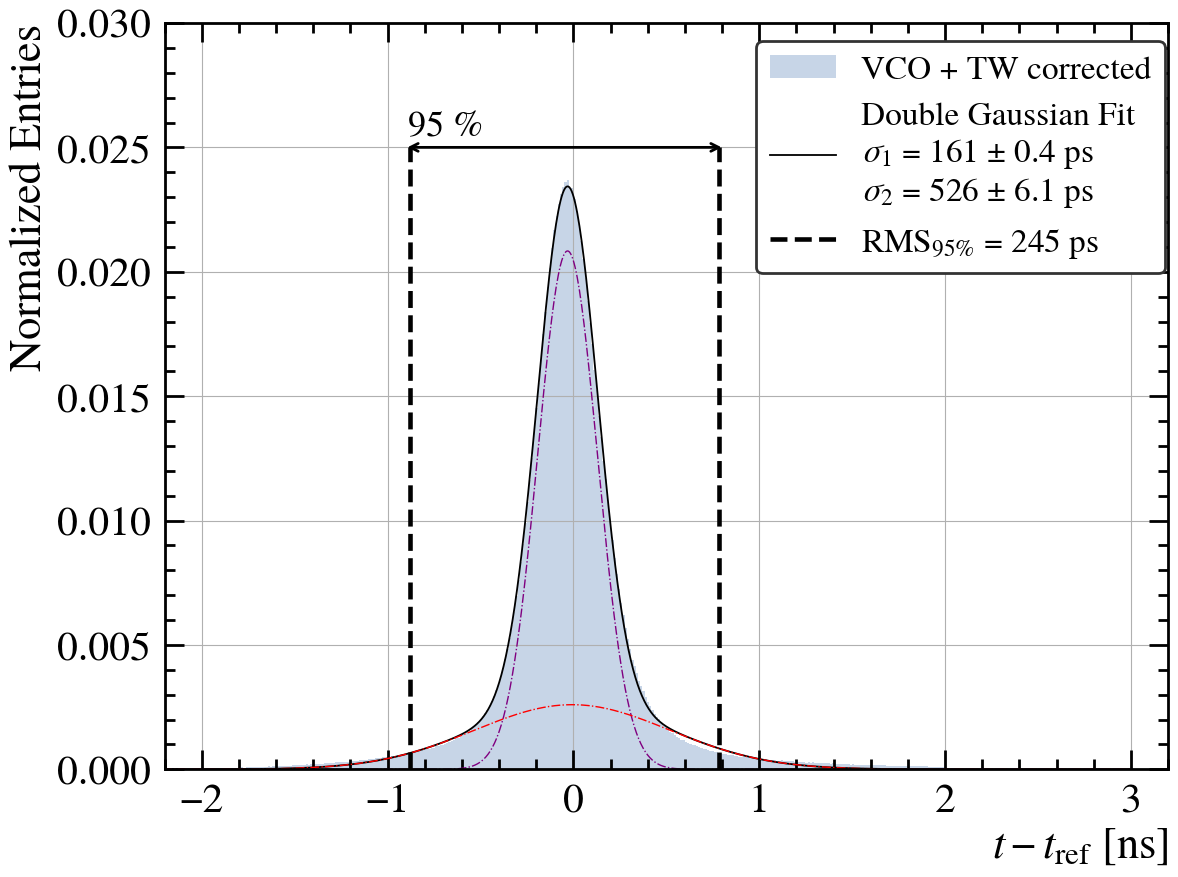}
    \caption{Time residual distribution after \gls{vco} and position-informed timewalk correction for all events, normalized to unit area. The sensor was reverse-biased at 65\,V with a charge threshold of $1\ke$. The results of a fit using the sum of two Gaussian functions with a common mean and different amplitudes and widths are superimposed. \label{fig:double_gaus}}
\end{figure}

Since there is a strong correlation between the generated charge and the intrapixel hit position of the tracks, a timewalk correction is derived in $1\times1\,\unit{\micro m}^2$ bins of the intrapixel area and applied globally to all pixels. This type of timewalk function is formulated according to \cref{eq:vco_corr}, but accounts for the track impact position in the 3D pixel and is therefore referred to as position-informed timewalk correction. In \cref{fig:double_gaus}, the density distribution of time residuals for all events is shown, corrected for the \gls{vco} variations and with the position-informed timewalk method. The distribution is symmetric around zero and has long tails, which are caused by particles passing through one of the electrodes. A sum of two Gaussian functions with a common mean and different widths denoted by $\sigma_1$ and $\sigma_2$ is fit to the distribution. The need for the second Gaussian function is attributed to the existence of roughly two separate regions within the pixel, where a worse resolution is obtained close to the electrodes. The effective width of this distribution is estimated as
\begin{equation}
\label{eq:double_gaus_sigma}
\sigma_\text{t} = \sqrt{f_1\cdot\sigma_1^2+(1-f_1)\cdot\sigma_2^2}\,,
\end{equation}
where $f_1$ is the fraction of the distribution described by the first Gaussian function.  
In \cref{fig:double_gaus}, the effective time resolution of the 3D detector, operating at 65\,V reverse bias and $1\ke$ threshold, is $241\pm2$\,ps. For reference, the RMS of the central $\SI{95}{\percent}$ of the same distribution results in a time resolution of $245\pm3$\,ps, compatible with the fit result. For this reason, in the following analysis of different bias voltages and thresholds, the time resolution is derived according to the RMS value of the central $\SI{95}{\percent}$ of events in the time residual distributions.

\begin{figure} 
    \centering
    \includegraphics[width=.47\textwidth]{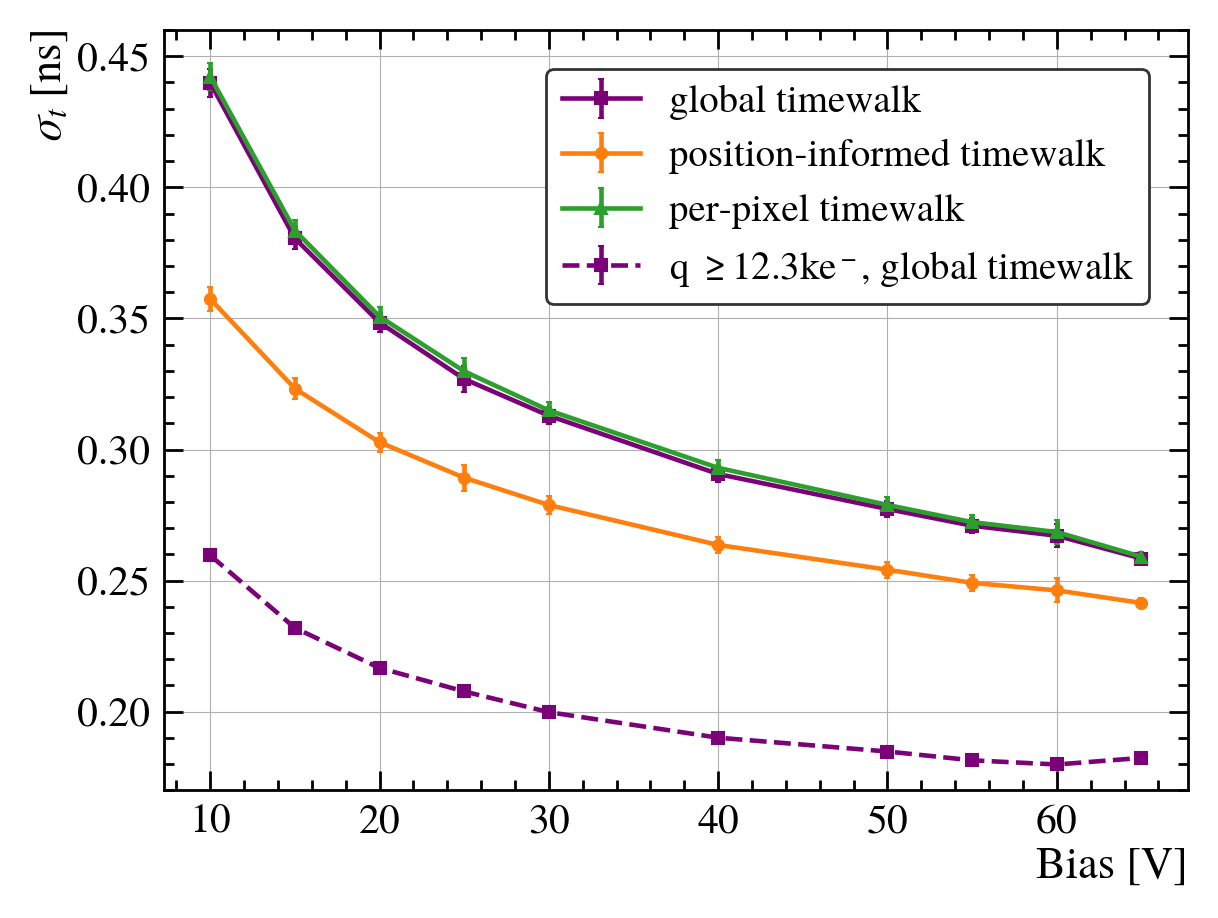}
    \quad
    \includegraphics[width=.48\textwidth]{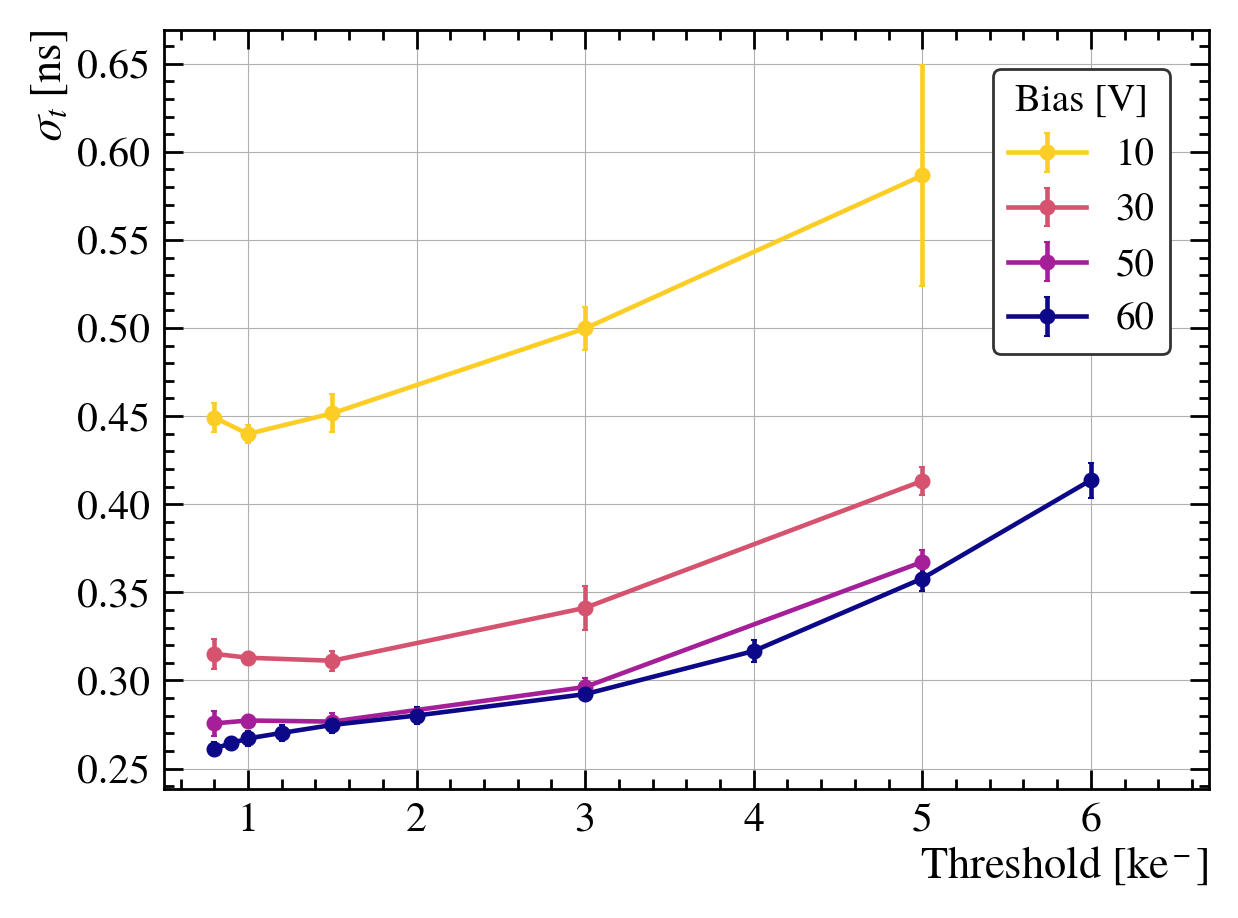}
    \caption{Left: Time resolution versus reverse bias voltage at $1\ke$ threshold as calculated after applying different types of timewalk corrections. Right: Time resolution versus charge threshold for different bias voltages, where measurements were corrected by global timewalk correction. The best performance occurs at the lowest threshold and highest bias voltage values. \label{fig:2D_scans}}
\end{figure}

A comparison of the time resolution given by the three different types of timewalk correction is shown in \cref{fig:2D_scans} (left). For a constant threshold value of $1\ke$ and an increasing reverse bias voltage, the time measurements are corrected by  a global charge-based, per-pixel or position-informed timewalk function. 
Interestingly, per-pixel timewalk corrections do not improve the time resolution compared to the global timewalk correction, whereas for some cases with limited statistics they may even result in worse performance. This can be explained by the effective removal of differences between pixel time offsets by the per-pixel \gls{vco} corrections. On the other hand, the position-informed timewalk corrections result in better time resolution as a result of less skewed residual distributions. The time resolution improves with increasing bias voltage due to stronger electric fields, resulting in faster drift of the charge carriers, with the diminishing improvement at higher voltages being consistent with the carrier drift velocity approaching saturation. The best time resolution in the 3D detector is obtained for the highest possible reverse bias that could be achieved before breakdown occurs, i.e. 65\,V. For this voltage, the time resolution is equal to $258\pm4$\,ps for the global timewalk correction and $245\pm3$\,ps for the position-informed timewalk correction, leading to a relative improvement of $\SI{6}{\percent}$.

\begin{figure}
    \centering
    \includegraphics[width=.47\textwidth]{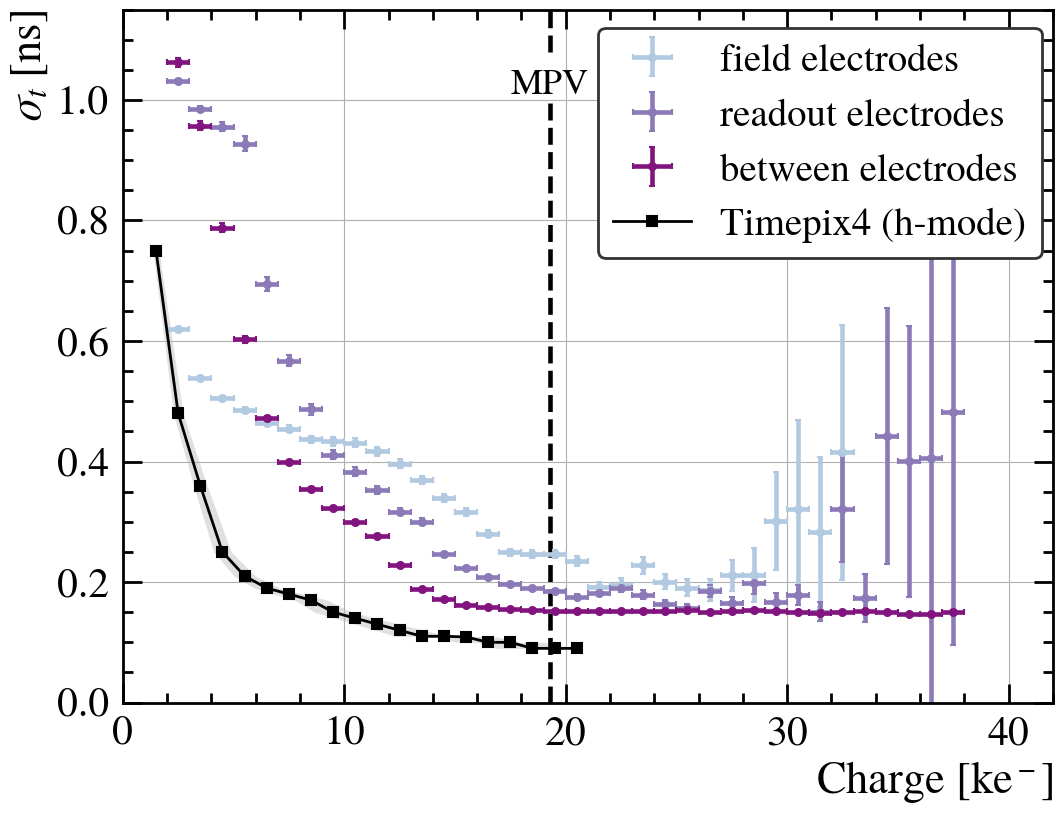}
    \quad
    \includegraphics[width=.465\textwidth]{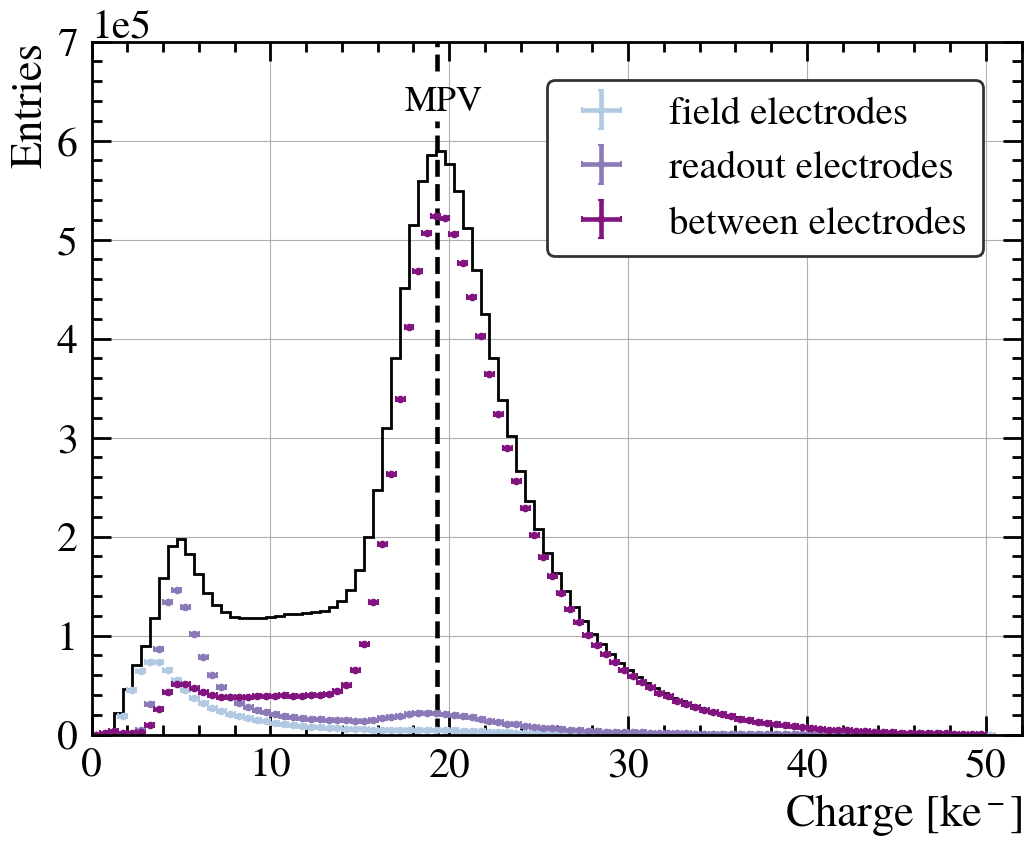}
    \caption{Left: Time resolution of the 3D detector as a function of the hit charge. The three colored lines correspond to events in the region of field electrodes, readout electrodes or between the electrodes. The black line shows the Timepix4 front-end time resolution for the hole collecting mode. Right: Hit charge distribution for all hits. The colored points show the scaled charge distributions for events in the region of field electrodes, readout electrodes or between the electrodes. The MPV charge for hits between the electrodes is drawn in both plots.\label{fig:resol_vs_q}}
\end{figure}

In order to study the impact of the discriminator threshold on the time resolution, a threshold scan was performed. In \cref{fig:2D_scans}~(right), the time resolution versus the charge threshold value is given for different reverse bias voltages, after the global timewalk correction was calculated and applied for each set of bias and threshold settings. The best performance of the 3D detector occurs at the lowest threshold value and the highest bias. With increasing threshold values, an increasing number of hits below the electrodes do not deposit enough charge to cross the threshold. In such cases, the time residual distributions are dominated by charged particles passing between the electrodes. Even with this effect, the time resolution at high threshold is still worse than at low threshold values. 
An estimate of the time resolution that could be achieved using only hits with high \gls{tot}, while the time measurement at a low threshold is maintained, has been obtained offline. By requiring a minimum \gls{tot} value equivalent to $12.29\pm0.96\ke$, events in the electrode regions are effectively rejected (see \cref{fig:resol_vs_q}, right). The results of this analysis are shown by the dashed line on \cref{fig:2D_scans}~(left), with a best time resolution of 180\,ps. However, this analysis discards $\SI{25}{\percent}$ of the reconstructed clusters, drastically reducing the detection efficiency of the 3D sensor.

\Cref{fig:resol_vs_q} (left) shows the charge-dependent time resolution of the 3D detector at 65\,V reverse bias and $1\ke$ threshold after applying the position-informed timewalk corrections. Three different curves are plotted to distinguish the time behavior of the 3D sensor for events in the regions of field electrodes, readout electrodes, or between electrodes, following the definition of them in \cref{fig:intrapixel}. These curves should be considered together with the charge distribution histogram in  \cref{fig:resol_vs_q}~(right).
In a charge window around the \gls{mpv} charge for events between the electrodes, $19.3\pm 0.5\ke$, the time resolution is $153\pm0.4$\,ps. Events in the same charge window, but with an impact position in the electrode regions instead, present a deteriorated time resolution of around 250\,ps. This is not a timewalk-related effect but instead indicates variations in the weighting potential~\cite{ramo} and the drift field. At the \gls{mpv} signal charge, the analog front-end jitter for the hole collecting mode (h\textsuperscript{+} mode) of the tested Timepix4 chip was measured to be $90\pm4$\,ps, as shown by the black curve obtained with test pulse injection~\cite{tpx4_resol}. By adding the \gls{tdc} and the analog front-end components in quadrature, the total contribution of Timepix4 to the time resolution is $108\pm3$\,ps. Therefore, the time resolution is limited by the 3D sensor itself, in contrast to the reported timing performance of the exact same 3D silicon sensor on a Timepix3 \gls{asic} in~\cite{3DonTPX3}, where the main limitation was imposed by the analog front-end jitter of Timepix3.
The time resolution is rather affected by the non-uniformity of the weighting potential in this \gls{3d-ddtc} configuration and variations in the drift field across the pixel.

\subsection{Small-angle studies} \label{subsec:low_angle}

\begin{figure}
    \centering
    \begin{minipage}[c]{.46\textwidth}
    \centering
    \includegraphics[width=\textwidth]{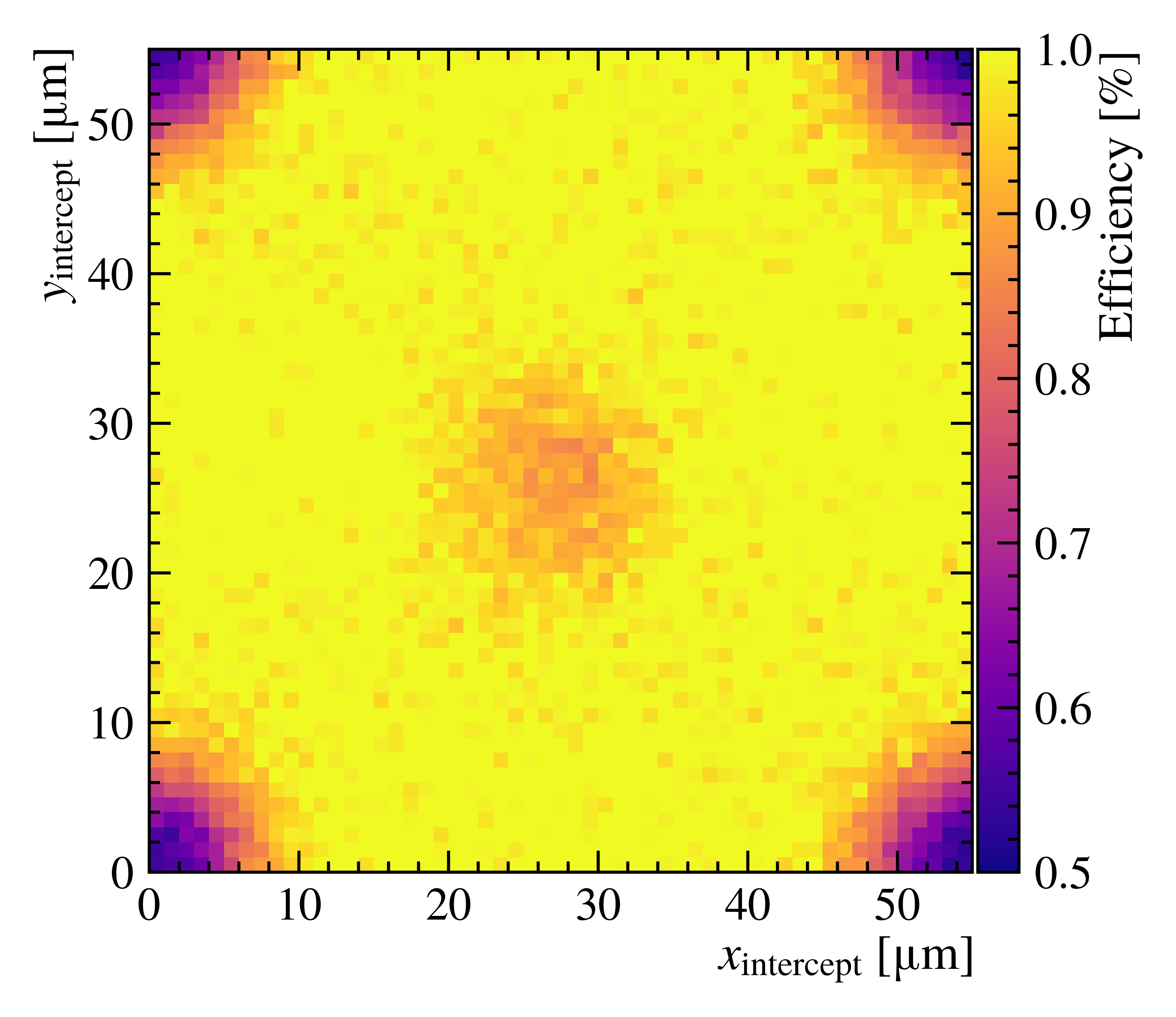}
    \end{minipage}
    \begin{minipage}[c]{.51\textwidth}
    \centering
    \includegraphics[width=\textwidth]{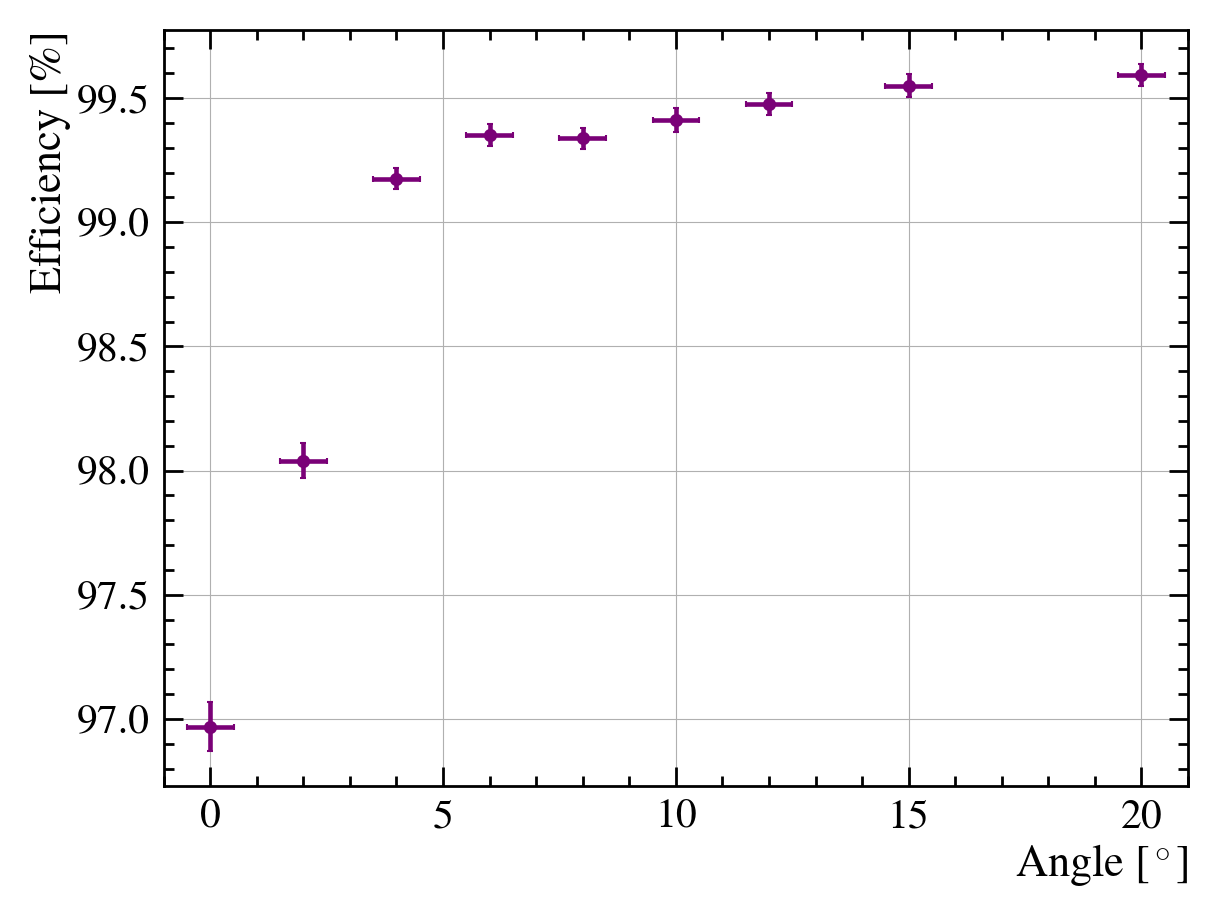}
    \end{minipage}

    \caption{Left: Intrapixel hit efficiency at normal incidence. Right:  Average hit detection efficiency as a function of the track-incidence angle $\phi$.\label{fig:eff}}
\end{figure}

In 3D sensors, the electrodes are essentially regions of inactive volume, so if most of a particle trajectory passes through them, it will not be detected. A common solution to recover the geometrical efficiency of 3D sensors is to place them at an angle with respect to the beam direction. In this way, a significant fraction of all charged-particle tracks pass through the active silicon material of the detector, where they deposit some detectable charge. In the tested \gls{3d-ddtc} sensor, the problem of insensitive electrode regions is mitigated by electrodes that do not fully penetrate the bulk. In the regions below the electrodes, the traversing particles interact and deposit some charge. Therefore, the hit efficiency of the 3D detector under study for perpendicularly incident tracks at a reverse bias voltage of 60\,V and at a threshold of $1\ke$ is $\SI{97}{\percent}$. According to \cref{fig:eff}~(left), the inefficiency mainly arises from particles going through the field electrodes, where charge sharing is more probable. As shown in \cref{fig:eff}~(right), a small rotation of the sensor, around $\phi=4^\circ$, is sufficient to reach a hit detection efficiency above $\SI{99}{\percent}$. 

\begin{figure}
    \centering
    \includegraphics[width=0.95\textwidth]{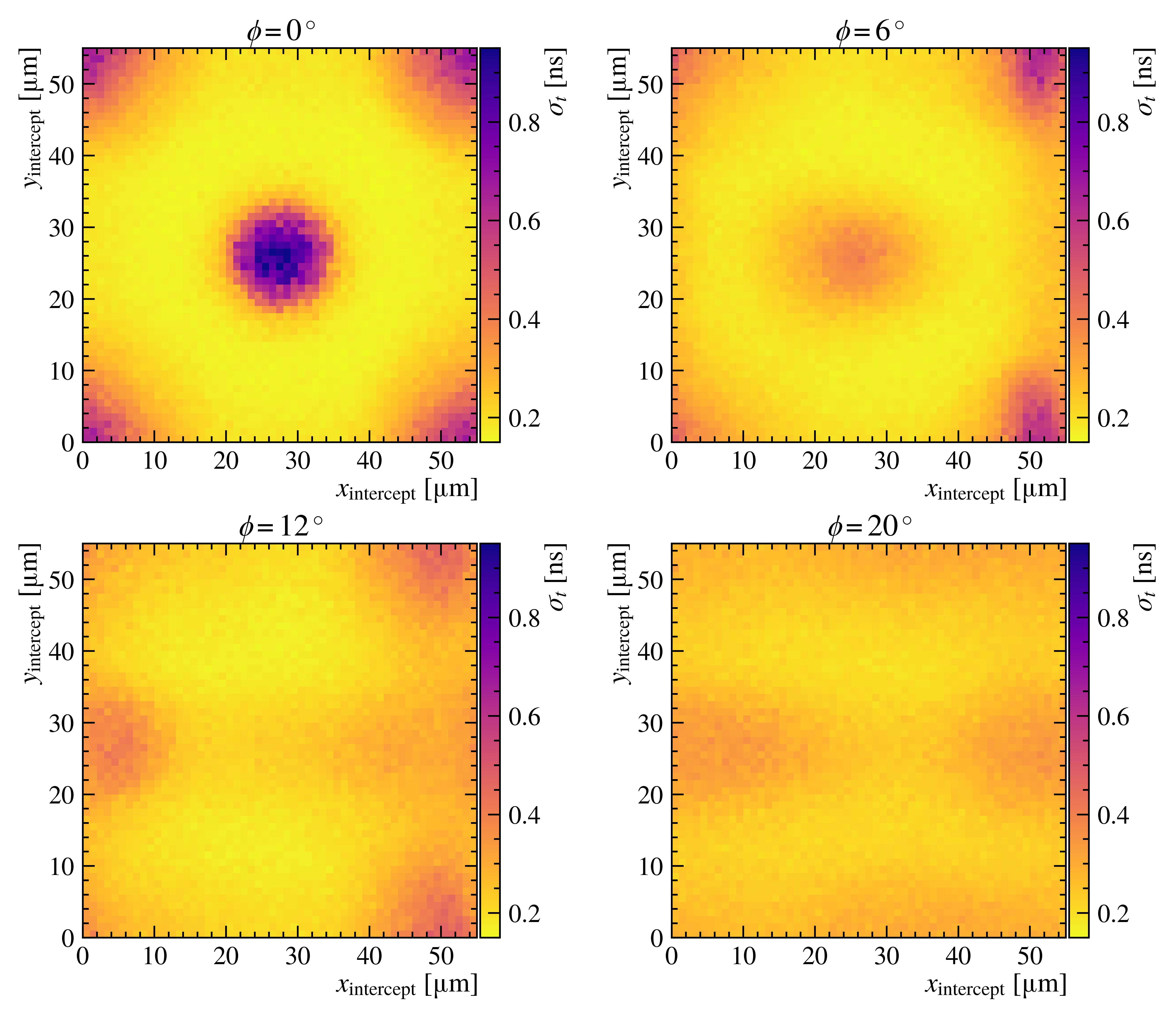}
    \caption{Time resolution of the 3D detector as a function of intrapixel position for different angles $\phi$. For increasing $\phi$ values, the center of the signal pillar seems to move. The distortion of the 3D pixel is due to the definition of the intrapixel x,y hit position in the range $[0\,\unit{\micro m},55\,\unit{\micro m}]$. 
    \label{fig:intra_resol_theta}}
\end{figure}

As discussed in \cref{subsec:perpendicular}, tracks with a hit in the electrode regions are associated with worse timing performance due to the small \gls{mpv} of deposited charge. Turning the sensor at a small angle also mitigates this problem and helps improve the time resolution across the entire pixel area. A qualitative illustration of how this works is presented in \cref{fig:intra_resol_theta}. The pixel is segmented into $1\times1\unit{\micro m}^2$ bins and the time resolution within the pixel area is shown for different angles $\phi$ (rotation around the $y$-axis). The time measurements are corrected by the position-informed timewalk, where for each angle and bin a set of timewalk parameters is determined. Although this type of timewalk correction is expected to minimize the impact of the electrodes on time resolution, their presence is clearly visible in the top-left plot of \cref{fig:intra_resol_theta}. As the 3D detector is rotated, the poor time resolution in the regions containing pillars gradually vanishes, and the time response of the 3D sensor becomes more uniform inside the pixel.

\begin{figure}
    \centering
    \includegraphics[width=.48\textwidth]{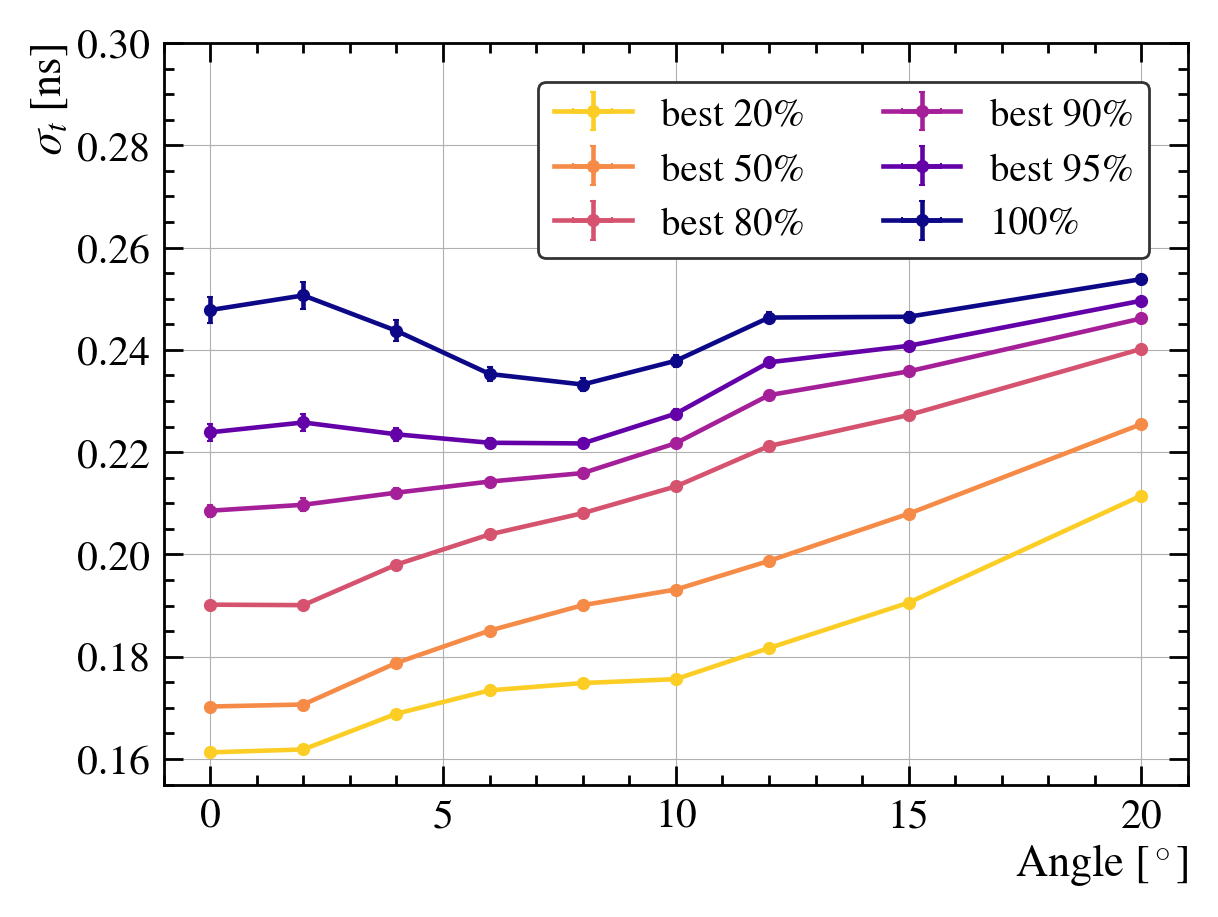}
    \caption{Time resolution of the 3D detector as a function of the rotation angle $\phi$. Each line considers a certain percentage of the pixel area. \label{fig:percents}}
\end{figure}

To quantify the influence of the sensor rotation angle on the uniformity of temporal performance, the time resolution for different best-performing percentiles of the total pixel area is determined, as shown in \cref{fig:percents}. For each run, the bias voltage was kept at 60\,V and the threshold value set at $1\ke$, while the angles were scanned in the range [\ang{0}, \ang{20}]. Although the cluster size increases for larger angles, only the time information from the first hit is used and corrected by the position-informed timewalk.
As shown in \cref{fig:percents}, even when considering tracks from the best $\SI{90}{\percent}$ of the pixel, the best timing performance corresponds to the normal incidence, at $\phi=0^\circ$. However, for $\SI{95}{\percent}$ or $\SI{100}{\percent}$ of the pixel area, the optimal angle is \ang{8}. For $\SI{100}{\percent}$ of the pixel area, the time resolution is equal to $248\pm3$\,ps at $\phi=0^\circ$ and $233\pm3$\,ps at $\phi=8^\circ$, marking $\SI{6}{\percent}$ improvement. While sensor rotation improves the overall time resolution of the detector, the timing performance of the best-performing pixel regions deteriorates.

 \subsection{Grazing-angle incidence}\label{subsec:grazing}
 
At grazing-angle incidence, the \gls{dut} stage is rotated around the $y$-axis, placing the 3D detector almost parallel to the beam direction. The grazing angle method enables the investigation of the timing behavior for different depths from the sensor surface. A detailed explanation of this method is presented in~\cite{grazing_method}. Incoming particles enter from the backside of the 3D sensor, traverse multiple adjacent pixels, and exit from the frontside, where the bump-bonded \gls{asic} is located. Each particle results in the formation of long clusters across multiple rows of the Timepix4 matrix. For a nominal rotation angle of the stage $\phi=88^\circ$, the mean cluster length is $111.9\pm0.8$ pixels. The charged particles cross each pixel at a different depth, with the average depth for the $i$\textsuperscript{th} pixel, with respect to the minimum Timepix4 row of the cluster, given by
\begin{equation}
\label{eq:depth_vs_theta}
d(i) = \frac{55\,\unit{\micro m} \times i}{\text{tan}\phi}\,
\end{equation}
where a depth of 0\um corresponds to the ASIC side, while a full depth of 300\um corresponds to the backside.

In the case of fully passing-through columns in a 3D sensor, the electric field is independent of depth, neglecting surface effects. However, in this \gls{3d-ddtc} variation with partially penetrating electrodes, a $z$-component (using the coordinate system defined in \cref{fig:3D_scheme}) of the electric field is introduced and becomes significant near the ends of the pillars. Three different $z$-segments of interest can be defined: (i) $z \leq 60$\um, where the field electrodes are absent; (ii) \mbox{$60\,\unit{\micro m}< z < 230\,\unit{\micro m}$}, which corresponds to a fully 3D pillar configuration; and (iii) $z \geq 230$\um, where the readout electrodes are absent. For these three $z$-segments, the hit charge \gls{mpv} has been measured to be $2.31\ke$ for $z \leq 60$\um, $2.28\ke$ for \mbox{$60\,\unit{\micro m}< z < 230\,\unit{\micro m}$} and $2.59\ke$ for $z \geq 230$\um. These values are lower than typical charge deposition in roughly 55\um of silicon, but the aforementioned charge calibration mismatch as well as an effective width of the columns above 10\um could explain the difference. Similar \glspl{mpv} for $z \leq 60$\um and  $z \geq 230$\um would be expected, since these segments are geometrically identical. However, a difference is observed, also for different bias voltages, and is attributed to the increased charge sharing in the $z \leq 60$\um segment.

\begin{figure}
    \centering
    \includegraphics[width=.485\textwidth]{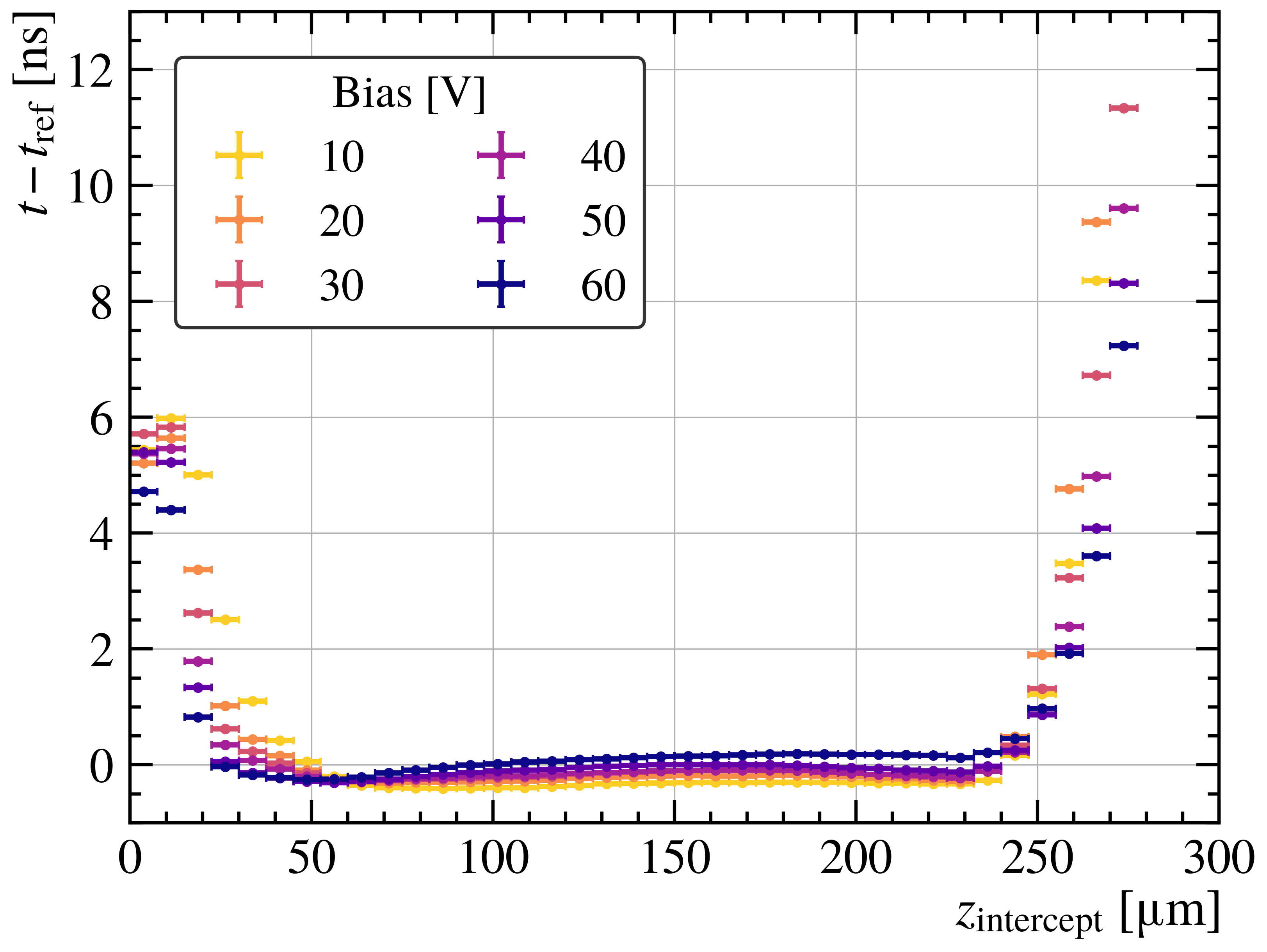}
    \quad
    \includegraphics[width=.48\textwidth]{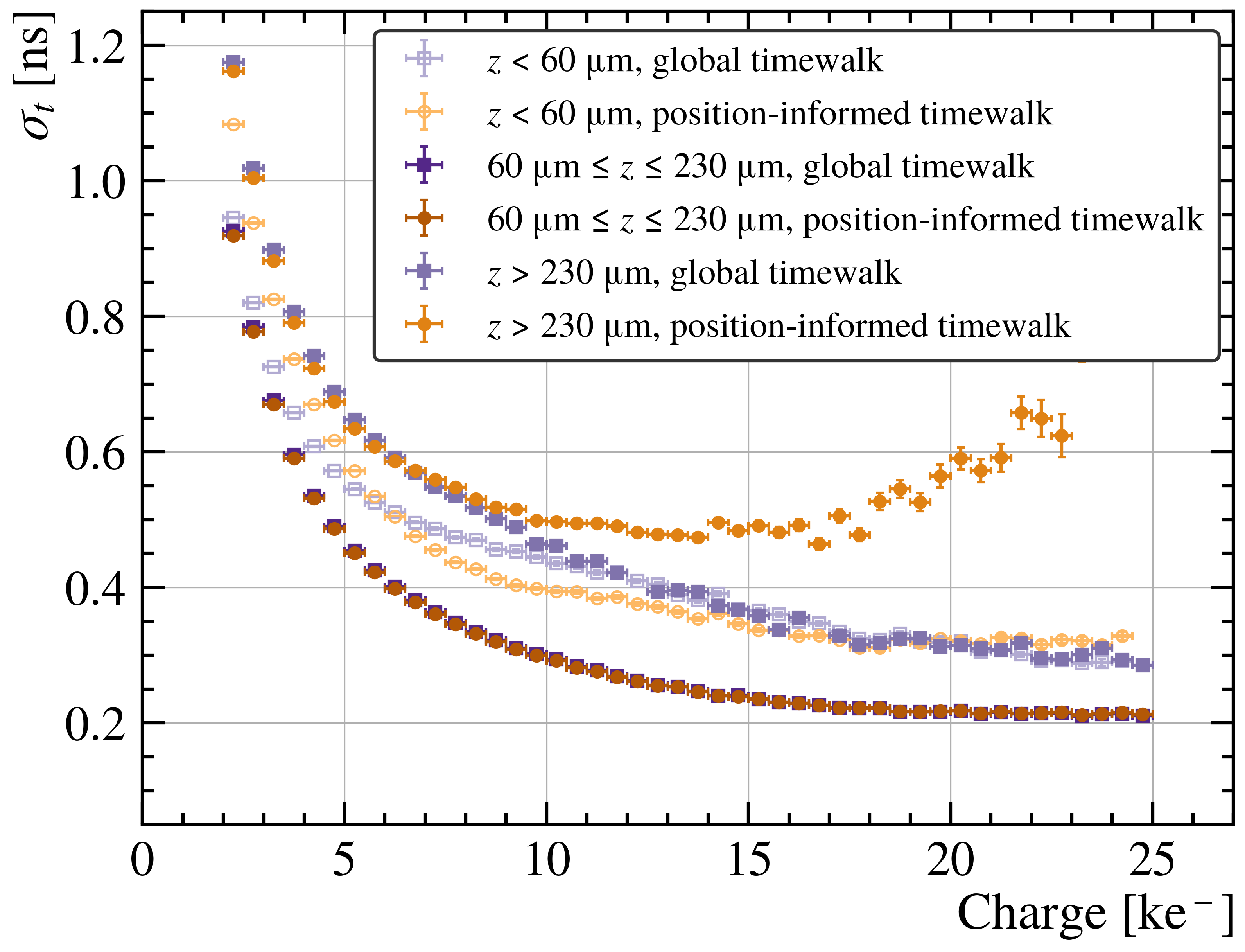}
    \caption{Left: Relative time delay versus sensor depth $z$ for different bias voltages. The relative delay values are corrected for \gls{vco} frequency and timewalk. Right: Time resolution as a function of the hit charge. Different colored curves correspond to different $z$-segments within the 3D sensor and different types of timewalk correction applied.\label{fig:time_grazing_angle}}
\end{figure}

At grazing-angle incidence, the charge is distributed relatively uniformly among the pixels in the cluster, and therefore no leading pixel can be selected for timing. The reference time is subtracted from the time measurement of each pixel, and \gls{vco} and timewalk corrections are applied. A global timewalk function, according to \cref{eq:tw_corr}, is calculated for this data. In addition, position-informed timewalk corrections are applied for different depths. The 300\um thickness of the 3D sensor is segmented into 50 bands of 6\um depth and, for each segment, a timewalk function $\Delta t_\text{tw}(q,z)$ is calculated. 
The corrected relative time delays versus the $z$ intercept are plotted for various bias voltages in \cref{fig:time_grazing_angle}~(left). An increase is observed for depths $z \leq 60$\um and especially $z \geq 230$\um, close to the surfaces of the 3D sensor. Although global timewalk parameters are used for this plot, the results do not improve when applying position-informed timewalk corrections. Since the \gls{mpv} measurements indicate that charge is still generated at shallow depths, the slow charge collection suggests a lower drift-field strength.
The absence of the opposite-type electrodes in these regions results in a low electric field~\cite{3d_simul_old}. The drift velocity of the eh pairs is small, and charges are partially collected by diffusion. 

On the right side of \cref{fig:time_grazing_angle}, the time resolution at 60\,V is shown as a function of the hit charge. Each curve corresponds to a specific $z$-segment and the two different types of timewalk corrections are applied. Overall, the time performance of the 3D detector for depths \mbox{$60\,\unit{\micro m}< z < 230\,\unit{\micro m}$} is notably better than for $z \leq 60$\um and $z \geq 230$\um. For all hits in the $z$-segment \mbox{$60\,\unit{\micro m}< z < 230\,\unit{\micro m}$}, the time resolution is equal to $720\pm6$\,ps. This plot supports the same conclusion as \cref{fig:resol_vs_q}~(left): hits in the electrode regions are produced at shallow depths and result in worse time resolution compared to hits between the electrodes, which are produced along the entire detector thickness.

\begin{figure}
    \centering
    \includegraphics[width=.47\textwidth]{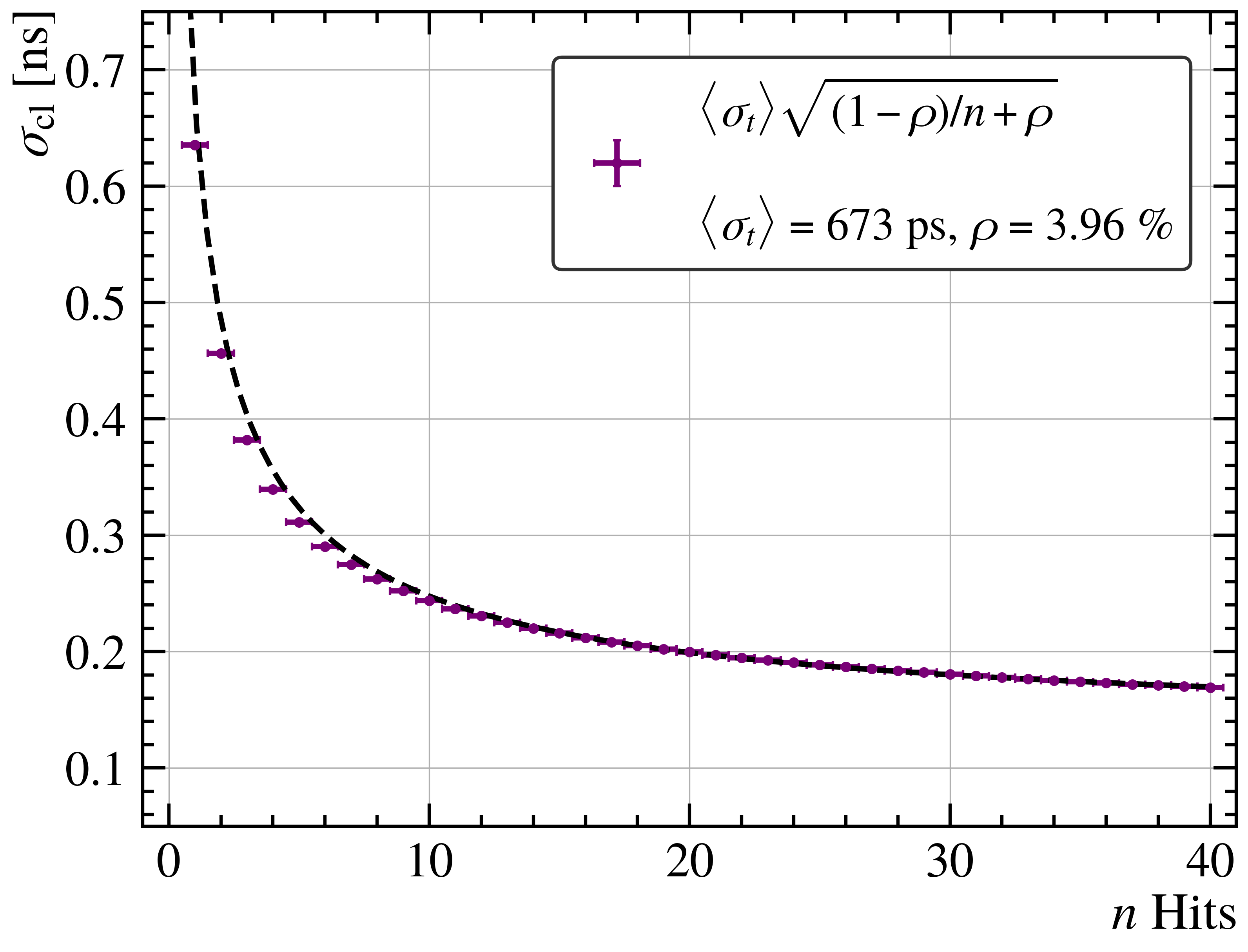}
    \caption{Cluster time resolution as a function of the number of hits used to calculate the cluster time.\label{fig:resol_vs_nhits}}
\end{figure}

The grazing angle method also offers the possibility to accurately timestamp passing charged particles by utilizing the time information provided by multiple pixels. A cluster time measurement can be extracted as the mean of the single-hit measurements, weighted by the charge dependent time resolution in \cref{fig:time_grazing_angle} (right). If $n$ single-hit time measurements from the $z$-segment of \mbox{$60\,\unit{\micro m}< z < 230\,\unit{\micro m}$} are used, and the index $i$ refers to individual pixel measurements, the cluster time $t_\text{cl}$ is given by
\begin{equation}
\label{eq:multi_hit_time}
t_\text{cl}(n) = \frac{\sum\limits_{i=0}^{n}{t_i/\sigma^2_t(q_i)}}{\sum\limits_{i=0}^{n}{1/\sigma^2_t(q_i)}}\,,
\end{equation}
where weighting by $1/\sigma^{2}_t$ minimizes the error of the cluster time estimator.
Considering uncorrelated single-hit time measurements with an average single-hit resolution $\langle\sigma_t\rangle$, the time resolution of the cluster time measurement should scale with $1/\sqrt{n}$. However, some correlation between the single-hit time measurements might exist, since they all refer to a single 40\,MHz clock and \glspl{vco} are shared in superpixels. This correlation is much stronger for perfectly even \gls{tdc} bins.
In reality, in Timepix4 \gls{vco} copies are not identical, as previously discussed, and variations in the \gls{tdc} bin size, phase offsets between them, and potential differences of the charge carrier's drift time act as a decorrelation mechanism for the single-hit time measurements. The dependence of the cluster time resolution on the number of hits, incorporating a correlation factor $\rho$, can be approximated by
\begin{equation}
\label{eq:multi_hit_resol}
\sigma_\text{cl}(n) = \langle\sigma_t\rangle\sqrt{\frac{1-\rho}{n}+\rho}\,.
\end{equation}
The cluster time resolution as a function of the number of hits used to calculate the weighted time is presented in \cref{fig:resol_vs_nhits}. Performing a fit using \cref{eq:multi_hit_resol}, a correlation factor of $\rho=3.96\si{\percent}$ is found, indicating that hits are mostly uncorrelated. Despite the small value of the correlation, it has a strong impact on the cluster time resolution for large cluster sizes. The time resolution for clusters of 40 hits is approximately 170\,ps, while for completely uncorrelated hits it is expected to be around 106\,ps. 
Further investigations of these correlations are of interest for future \gls{4d} tracking \glspl{asic} as well as ultrafast timing detectors. 

\section{Conclusions}  \label{sec:conclusions}

A 3D silicon sensor read out by a Timepix4 \gls{asic} was evaluated in terms of its spatial and time resolution.
The 3D detector was tested with a 180 GeV/c mixed hadron beam using the Timepix4 telescope at the CERN SPS H8 beamline.
The data analysis encompassed $\eta$ corrections as well as corrections to compensate for the \gls{vco} frequency variations and timewalk.

The 3D detector has a time resolution of $245\pm3$\,ps at a 65 V reverse bias and $1\ke$ threshold. Charged particles with an impact position between the electrodes display a time resolution of $153\pm0.4$\,ps, corresponding to a hit-charge \gls{mpv} roughly of $19.3\ke$. Even for the best reported results, the dominant contribution to the time resolution is imposed by the 3D sensor, since the time resolution of Timepix4 \gls{asic} in the hole-collecting mode was measured to be below 110\,ps.
 
Rotating the \gls{dut} with respect to the beam direction benefits both the hit detection efficiency, as well as the temporal and spatial resolution of the 3D detector. At an optimal angle of \ang{8}, the time resolution improves by $\SI{6}{\percent}$, while the hit efficiency increases from $\SI{97}{\percent}$ to $\SI{99.3}{\percent}$. At the same angle, the best spatial resolution of approximately 7\um is achieved for MIPs traversing the sensor, using a threshold of 0.8\ke and $\eta$-corrected charge-weighted clusters. This angle is found to give the optimal spatial resolution due to increased charge sharing between neighbouring pixels, corresponding to a charged-particle displacement across approximately one full pixel pitch. At the same time, most of the deposited charge remains confined to a single pixel, preserving large signal amplitudes that benefit the temporal resolution.

A further rotation of the 3D detector to a grazing angle of $\phi=88^\circ$ revealed worse timing performance at depths where opposite-type electrodes are absent. A brief study of the possibility of achieving better time resolution by combining single-hit time measurements was presented. The time resolution at grazing angle incidence is equal to 720\,ps for single-hit time measurements and improves to 170\,ps for a cluster size of 40, with $\SI{4}{\percent}$ correlation. A thorough understanding of the origin of these correlations could help mitigate their impact and potentially achieve a time resolution as good as the lowest achievable value for this \gls{asic}. 

The limitations in the measured time resolution can be attributed to several factors.
An important limitation arises from the binary-like timing behavior observed, where regions between the pillars exhibit good time resolution, while the pillar regions show significantly worse performance. Reducing the electrode diameter would improve the ratio between these regions, although its impact on the weighting-potential uniformity should also be studied.
In this context, alternative electrode configurations with multiple readout columns per pixel, such as the \gls{2e} geometry~\cite{1e2e}, one of the best candidates for the \gls{velo} Upgrade~II, could provide a more uniform weighting field and improve the timing performance. However, additional electrodes increase the pixel capacitance and therefore the electronics jitter.
The tested sensor is read out via p-type columns and requires the Timepix4 \gls{asic} to operate in hole-collecting mode, while Timepix4 provides its best time resolution for electron-collecting sensors~\cite{tpx4}. 
Future 3D sensors with n-type readout bonded to Timepix4 are expected to achieve improved results.
Finally, in the tested \gls{3d-ddtc} design, the electric field depends on the sensor depth, introducing a spread in the drift velocity of the charge carriers. 
A 3D design with fully etched-through electrodes could reduce this spread and lead to a more uniform drift field, although this may come at the cost of increased capacitance and decreased hit efficiency. Overall, future studies could play an important role in optimizing 3D sensor design by balancing the aforementioned effects.

\acknowledgments
We express our gratitude to the CERN SPS accelerator team and beamline physicists for ensuring great performance of the SPS. 
We thank the LHCb testbeam coordinator, Loris Martinazzoli, for the support and coordination of the testbeam campaigns.
We acknowledge the engineers who designed the Timepix4 ASIC.
We gratefully acknowledge the support from the following national agencies: the Netherlands Organisation for Scientific Research (NWO), in particular this publication is part of the project FASTER with file number OCENW.XL21.XL21.076 and the Science and Technology Facilities Council (United Kingdom) including grant reference ST/@004305/1.

\bibliographystyle{JHEP}
\bibliography{biblio.bib}

\end{document}